\documentclass[useAMS,usenatbib]{mn2e}

\usepackage[dvips]{graphicx}

\title[The initial-final mass relationship of white dwarfs revisited]
      {The initial-final mass relationship of white dwarfs revisited: 
       effect on the luminosity function and mass distribution}
\author[S. Catal\'an et al.]
       {S. Catal\'an$^{1,2}$\thanks{E-mail: catalan@ieec.uab.es}, 
        J. Isern$^{1,2}$, 
        E. Garc\'\i a--Berro$^{1,3}$ and 
        I. Ribas$^{1,2}$\\
$^{1}$Institut d'Estudis Espacials de Catalunya, 
      c/ Gran Capit\`{a} 2--4, 08034 Barcelona, Spain\\
$^{2}$Institut de Ci\`encies de l'Espai, CSIC, 
      Facultat de Ci\`encies, Campus UAB, 08193 Bellaterra, Spain\\
$^{3}$Departament de F\'isica Aplicada, 
      Escola Polit\`ecnica Superior de Castelldefels, 
      Universitat Polit\`ecnica de Catalunya,\\
      Avda. del Canal Ol\'\i mpic s/n, 08860 Castelldefels, Spain}

\begin{document}

\date{\today}

\maketitle

\label{firstpage}

\begin{abstract} 
The initial-final mass relationship connects the mass of a white dwarf with the 
mass of its progenitor in the main-sequence.  Although this function is of 
fundamental importance to several fields in modern astrophysics, it is not well 
constrained either from the theoretical or the observational points of view.  
In this work we revise the present semi-empirical initial-final mass 
relationship by re-evaluating the available data.  The distribution obtained 
from grouping all our results presents a considerable dispersion, which is 
larger than the uncertainties.  We have carried out a weighted least-squares 
linear fit of these data and a careful analysis to give some clues on the 
dependence of this relationship on some parameters such as metallicity or 
rotation. The semi-empirical initial-final mass relationship arising from our 
study covers the range of initial masses from 1.0 to $6.5~M_{\sun}$, including 
in this way the low-mass domain, poorly studied until recently.  Finally, we 
have also performed a test of the initial-final mass relationship by studying 
its effect on the luminosity function and on the mass distribution of white 
dwarfs. This was done by using different initial-final mass relationships from 
the literature, including the expression derived in this work, and comparing 
the results obtained with the observational data from the Palomar Green Survey 
and the Sloan Digital Sky Survey (SDSS).  We find that the semi-empirical 
initial-final mass relationship derived here gives results in good agreement 
with the observational data, especially in the case of the white dwarf mass 
distribution.
\end{abstract}


\section{Introduction}

The initial-final mass relationship of white dwarfs links the mass of a white 
dwarf with that of its progenitor in the main-sequence. This function is of 
paramount importance for several aspects of modern astrophysiscs such as the 
determination of the ages of globular clusters and their distances, the study 
of the chemical evolution of galaxies, and also to understand the properties of 
the Galactic population of white dwarfs. However, we still do not have an 
accurate measurement of this relationship and, consequently, more efforts are 
needed from both the theoretical and the observational perspectives to improve 
it.

\citet{wei77} carried out the first attempt to empirically map this 
relationship, and provided also a recent revision \citep{wei00}. Although many 
improvements have been achieved in these 30 years, there are still some pieces 
missing in the puzzle. For instance, the dependence of this function on 
different parameters is still not clear (e.g.~metallicity, magnetic field, 
angular momentum). On the other hand, numerous works have dealt with the 
calculation of a theoretical initial-final mass relationship 
\citep{dom99,mar01}, but the differences in their evolutionary codes, such as 
the treatment of convection, the value of the assumed critical mass, which is 
the maximum mass of a white dwarf progenitor, or the mass loss prescriptions 
used lead to very different results.  The main differences between the 
different theoretical approaches to the initial-final mass relationship have 
been extensively discussed in \citet{wei00}.

From an observational perspective, most efforts up to now have focused on the 
observation of white dwarfs in open clusters, since this allows to infer the 
total age and the original metallicity of white dwarfs belonging to the cluster 
\citep{wil04,kal05}.  Open clusters have made possible the derivation of a 
semi-empirical initial-final mass relationship using more than 50 white dwarfs, 
although only covering the initial mass range between $2.5$ and $7.0~M_{\sun}$ 
because stellar clusters are relatively young and, hence, the white dwarf 
progenitors in these clusters are generally massive. The recent study of 
\citet{kal08} based in old open clusters (NGC 7789, NGC 6819 and NGC6791) has 
extended this mass range to smaller masses.  A parallel attempt to cover the 
low-mass range of the initial-final mass relationship has been carried out by 
\citet{cat08}. This was the first study of this relationship based in common 
proper motion pairs. The stars studied in \citet{cat08} are at shorter 
distances in comparison with star clusters and this allows a better 
spectroscopic study of both members of the pair, obtaining their stellar 
parameters with accuracy. At the same time, the study of these pairs enables a 
wide age and metallicity coverage of the initial-final mass relationship.

The aim of this work is to perform a revision of the initial-final mass 
relationship taking into account our recent results from studying white dwarfs 
in common proper motion pairs \citep{cat08} and the available data that are 
currently being used to define the initial-final mass relationship. The paper 
is organized as follows. In \S2 we present the analysis of the data that we 
will use to define the semi-empirical initial-final mass relationship. Section 
3 is devoted to the detailed analysis of the semi-empirical initial-final mass 
relationship derived in this paper and to give some clues on its dependence on 
some parameters, such as metallicity.  In \S4 and \S5 we compute the luminosity 
function and the mass distribution of white dwarfs considering different 
initial-final mass relationships and compare our results with the available 
observational data.  Finally in \S6 we summarize our main results and we draw 
our conclusions.

\section{Analysis of current available data}

\begin{table*}
\caption{Results from  re-evaluating the available  data.}
\begin{center}
\footnotesize{
\begin{tabular}{lcccccccc}
\hline      
\hline      
\noalign{\smallskip}
WD  & $T_{\rm eff}$ (K) & $\log g$ (dex)& $M_{\rm f}$ ($M_{\sun}$) & $t_{\rm cool}$ (Gyr) & $t_{\rm prog}$ (Gyr) & $M_{\rm i}$ ($M_{\sun}$) & $Z$ \\
\hline      
\noalign{\smallskip}
NGC 2099 (M37)& & & & & & & $0.011$\\
\noalign{\smallskip}
WD2 & 19900$\pm$900 & 8.11$\pm$0.16 & 0.69$\pm$0.07 & 0.093$\pm$0.019 & 0.56$\pm$0.07 & 2.72$_{-0.10}^{+0.12}$&\\ 
WD3 & 18300$\pm$900 & 8.23$\pm$0.21 & 0.76$\pm$0.09 & 0.152$\pm$0.034 & 0.50$\pm$0.07 & 2.82$_{-0.13}^{+0.16}$&\\
WD4 & 16900$\pm$1100& 8.40$\pm$0.26 & 0.86$\pm$0.12 & 0.259$\pm$0.072 & 0.39$\pm$0.10 & 3.06$_{-0.22}^{+0.32}$&\\
WD5 & 18300$\pm$1000& 8.33$\pm$0.22 & 0.82$\pm$0.11 & 0.192$\pm$0.078 & 0.46$\pm$0.10 & 2.90$_{-0.19}^{+0.26}$&\\
WD7 & 17800$\pm$1400& 8.42$\pm$0.32 & 0.88$\pm$0.14 & 0.219$\pm$0.082 & 0.43$\pm$0.10 & 2.96$_{-0.21}^{+0.30}$&\\
WD9 & 15300$\pm$400 & 8.00$\pm$0.08 & 0.61$\pm$0.03 & 0.182$\pm$0.015 & 0.47$\pm$0.07 & 2.88$_{-0.13}^{+0.16}$&\\
WD10& 19300$\pm$400 & 8.20$\pm$0.07 & 0.74$\pm$0.03 & 0.120$\pm$0.010 & 0.53$\pm$0.06 & 2.76$_{-0.11}^{+0.13}$&\\
WD11& 23000$\pm$600 & 8.54$\pm$0.10 & 0.98$\pm$0.04 & 0.136$\pm$0.014 & 0.51$\pm$0.07 & 2.80$_{-0.11}^{+0.13}$&\\
WD12& 13300$\pm$1000& 7.91$\pm$0.12 & 0.56$\pm$0.05 & 0.239$\pm$0.043 & 0.41$\pm$0.08 & 3.01$_{-0.17}^{+0.23}$&\\
WD13& 18200$\pm$400 & 8.27$\pm$0.08 & 0.78$\pm$0.03 & 0.167$\pm$0.016 & 0.48$\pm$0.07 & 2.86$_{-0.12}^{+0.15}$&\\
WD14& 11400$\pm$200 & 7.73$\pm$0.16 & 0.45$\pm$0.07 & 0.282$\pm$0.039 & 0.37$\pm$0.07 & 3.12$_{-0.19}^{+0.26}$&\\
WD16& 13100$\pm$500 & 8.34$\pm$0.10 & 0.82$\pm$0.05 & 0.480$\pm$0.062 & 0.17$\pm$0.09 & 4.10$_{-0.57}^{+1.44}$&\\
\hline      
\noalign{\smallskip}
NGC 2168 (M35)& & & & & & & $0.012$\\
\noalign{\smallskip}
LAWDS 1 & 32400$\pm$512 & 8.40$\pm$0.12 & 0.89$\pm$0.06 & 0.023$\pm$0.006   & 0.127$\pm$0.060   & 4.61$_{-0.64}^{+1.36}$  &\\
LAWDS 2 & 32700$\pm$603 & 8.34$\pm$0.08 & 0.85$\pm$0.04 & 0.017$\pm$0.003   & 0.133$\pm$0.060   & 4.53$_{-0.60}^{+1.21}$  &\\
LAWDS 5 & 52600$\pm$1160& 8.24$\pm$0.09 & 0.82$\pm$0.04 & 0.0022$\pm$0.0001 & 0.148$\pm$0.060 & 4.35$_{-0.52}^{+0.98}$&\\
LAWDS 6 & 55200$\pm$897 & 8.28$\pm$0.06 & 0.84$\pm$0.03 & 0.0020$\pm$0.0001 & 0.148$\pm$0.060 &  4.35$_{-0.52}^{+0.98}$  &\\
LAWDS 15& 29900$\pm$318 & 8.48$\pm$0.06 & 0.94$\pm$0.03 & 0.046$\pm$0.005   & 0.104$\pm$0.060   & 4.99$_{-0.81}^{+2.25}$  &\\
LAWDS 22& 54400$\pm$1203& 8.04$\pm$0.12 & 0.72$\pm$0.04 & 0.0025$\pm$0.0002 & 0.147$\pm$0.060 & 4.35$_{-0.52}^{+0.98}$&\\
LAWDS 27& 30500$\pm$397 & 8.52$\pm$0.06 & 0.98$\pm$0.03 & 0.048$\pm$0.004   & 0.102$\pm$0.060   & 5.03$_{-0.83}^{+2.37}$  &\\
\hline
\noalign{\smallskip}
NGC 3532 & & & & & & & $0.019$\\
\noalign{\smallskip}
3532-WD1 & 28000$\pm$2000 & 8.45$\pm$0.45 & 0.92$\pm$0.20 & 0.055$\pm$0.036 & 0.245$\pm$0.154 & 3.70$_{-0.55}^{+1.65}$ &\\
3532-WD5 & 28500$\pm$2000 & 7.8$\pm$0.3   & 0.55$\pm$0.11 & 0.013$\pm$0.002 & 0.287$\pm$0.150 & 3.52$_{-0.46}^{+1.05}$ &\\
3532-WD6 & 28500$\pm$3000 & 8.5$\pm$0.5   & 0.96$\pm$0.23 & 0.060$\pm$0.046 & 0.240$\pm$0.157 & 3.73$_{-0.57}^{+1.80}$ &\\
3532-WD8 & 23367$\pm$1065 & 7.71$\pm$0.15 & 0.48$\pm$0.05 & 0.023$\pm$0.003 & 0.277$\pm$0.150 & 3.56$_{-0.48}^{+1.14}$&\\
3532-WD9 & 29800$\pm$616  & 7.83$\pm$0.23 & 0.56$\pm$0.08 & 0.012$\pm$0.001 & 0.288$\pm$0.150 & 3.51$_{-0.46}^{+1.03}$ &\\
3532-WD10 & 19267$\pm$974  & 8.14$\pm$0.27 & 0.71$\pm$0.11 & 0.112$\pm$0.030 & 0.188$\pm$0.153 & 4.05$_{-0.73}^{+3.53}$&\\ 
\hline
\noalign{\smallskip}
Praesepe & & & & & & & $0.027$\\
\noalign{\smallskip}
WD0836$+$201 & 16629$\pm$350   & 8.01$\pm$0.05  & 0.62$\pm$0.02 & 0.144$\pm$0.009 & 0.481$\pm$0.051 & 2.97$_{-0.10}^{+0.11}$&\\
WD0836$+$199 & 14060$\pm$630   & 8.34$\pm$0.06  & 0.82$\pm$0.03 & 0.39$\pm$0.04   & 0.235$\pm$0.064 & 3.76$_{-0.28}^{+0.45}$&\\
WD0837$+$199 & 17098$\pm$350   & 8.32$\pm$0.05  & 0.81$\pm$0.03 & 0.22$\pm$0.02   & 0.405$\pm$0.054 & 3.15$_{-0.13}^{+0.15}$&\\
WD0840$+$200 & 14178$\pm$350   & 8.23$\pm$0.05  & 0.75$\pm$0.02 & 0.31$\pm$0.02   & 0.315$\pm$0.054 & 3.42$_{-0.17}^{+0.21}$&\\
WD0836$+$197 & 21949$\pm$350   & 8.45$\pm$0.05  & 0.91$\pm$0.03 & 0.13$\pm$0.01   & 0.495$\pm$0.051 & 2.94$_{-0.09}^{+0.11}$&\\
WD0837$+$185 & 14748$\pm$400   & 8.24$\pm$0.055 & 0.76$\pm$0.02 & 0.288$\pm$0.018 & 0.337$\pm$0.053 & 3.35$_{-0.16}^{+0.20}$&\\
WD0837$+$218 & 16833$\pm$254   & 8.39$\pm$0.03  & 0.85$\pm$0.01 & 0.251$\pm$0.009 & 0.374$\pm$0.051 & 3.23$_{-0.13}^{+0.16}$&\\
WD0833$+$194 & 14999$\pm$233   & 8.18$\pm$0.035 & 0.72$\pm$0.01 & 0.246$\pm$0.009 & 0.379$\pm$0.051 & 3.22$_{-0.13}^{+0.16}$&\\
WD0840$+$190 & 14765$\pm$270   & 8.21$\pm$0.03  & 0.74$\pm$0.01 & 0.27$\pm$0.01   & 0.355$\pm$0.051 & 3.29$_{-0.14}^{+0.17}$&\\
WD0840$+$205 & 14527$\pm$282   & 8.24$\pm$0.04  & 0.76$\pm$0.02 & 0.30$\pm$0.015  & 0.325$\pm$0.052 & 3.39$_{-0.16}^{+0.20}$&\\
WD0843$+$184 & 14498$\pm$202   & 8.22$\pm$0.04  & 0.75$\pm$0.02 & 0.295$\pm$0.012 & 0.330$\pm$0.051 & 3.35$_{-0.15}^{+0.19}$&\\
\hline
\noalign{\smallskip}
Hyades & & & & & & & $0.027$\\
\noalign{\smallskip}
WD0352$+$098 & 14770$\pm$350 & 8.16$\pm$0.05 & 0.71$\pm$0.02 & 0.25$\pm$0.01   & 0.375$\pm$0.051 & 3.23$_{-0.13}^{+0.16}$ &\\
WD0406$+$169 & 15180$\pm$350 & 8.30$\pm$0.05 & 0.79$\pm$0.02 & 0.29$\pm$0.02   & 0.335$\pm$0.054 & 3.35$_{-0.16}^{+0.20}$ &\\
WD0421$+$162 & 19570$\pm$350 & 8.09$\pm$0.05 & 0.68$\pm$0.02 & 0.096$\pm$0.006 & 0.529$\pm$0.050 & 2.88$_{-0.09}^{+0.10}$ &\\
WD0425$+$168 & 24420$\pm$350 & 8.11$\pm$0.05 & 0.70$\pm$0.02 & 0.038$\pm$0.003 & 0.587$\pm$0.050 & 2.78$_{-0.08}^{+0.09}$ &\\
WD0431$+$125 & 21340$\pm$350 & 8.04$\pm$0.05 & 0.65$\pm$0.02 & 0.060$\pm$0.004 & 0.565$\pm$0.050 & 2.82$_{-0.08}^{+0.09}$ &\\
WD0438$+$108 & 27390$\pm$350 & 8.07$\pm$0.05 & 0.68$\pm$0.02 & 0.018$\pm$0.001 & 0.607$\pm$0.050 & 2.75$_{-0.08}^{+0.08}$ &\\
WD0437$+$138 & 15335$\pm$350 & 8.26$\pm$0.05 & 0.77$\pm$0.02 & 0.26$\pm$0.01   & 0.365$\pm$0.051 & 3.26$_{-0.14}^{+0.17}$ &\\
\noalign{\smallskip}
\hline
\hline
\end{tabular}}
\label{tab:mif}
\end{center}
\end{table*}
\setcounter{table}{0}

\begin{table*}
\caption{Results    from     re-evaluating    the    available    data
         (continued). White  dwarfs in common proper  motion pairs are
         listed in the last section of this table (CPMPs)}
\begin{center}
\small{
\begin{tabular}{lcccccccc}
\hline
\hline
\noalign{\smallskip}
WD  & $T_{\rm eff}$ (K) & $\log g$ (dex)& $M_{\rm f}$ ($M_{\sun}$) & $t_{\rm cool}$ (Gyr) & $t_{\rm prog}$ (Gyr) & $M_{\rm i}$ ($M_{\sun}$) & $Z$ \\
\noalign{\smallskip}
\hline
\noalign{\smallskip}
NGC 2516 & & & & & & & $0.02$\\
\noalign{\smallskip}
2516-WD1 & 28170$\pm$310 & 8.48$\pm$0.17 & 0.95$\pm$0.08 & 0.060$\pm$0.012 & 0.081$\pm$0.012 & 5.54$_{-0.29}^{+0.39}$ &\\
2516-WD2 & 34200$\pm$610 & 8.60$\pm$0.11 & 1.01$\pm$0.04 & 0.035$\pm$0.008 & 0.106$\pm$0.008 & 5.03$_{-0.13}^{+0.14}$ &\\
2516-WD3 & 26870$\pm$330 & 8.55$\pm$0.07 & 0.99$\pm$0.03 & 0.082$\pm$0.005 & 0.059$\pm$0.005 & 6.44$_{-0.29}^{+0.32}$ &\\
2516-WD5 & 30760$\pm$420 & 8.70$\pm$0.12 & 1.07$\pm$0.05 & 0.074$\pm$0.014 & 0.067$\pm$0.014 & 6.01$_{-0.46}^{+0.69}$ &\\
\hline
\noalign{\smallskip}
NGC 6791 &  &  &  &  & & & $0.040$ \\
\noalign{\smallskip}
WD7 & 14800$\pm$300 & 7.91$\pm$0.06 & 0.56$\pm$0.02 & 0.17$\pm$0.01   & 8.33$\pm$0.85   & 1.086$_{-0.038}^{+0.045}$ &\\
WD8 & 18200$\pm$300 & 7.73$\pm$0.06 & 0.47$\pm$0.02 & 0.063$\pm$0.005 & 8.437$\pm$0.850 & 1.081$_{-0.037}^{+0.044}$ &\\
\hline
\noalign{\smallskip}
NGC 7789 &  &  &  &  & & & $0.014$\\
\noalign{\smallskip}
WD5 & 31200$\pm$200 & 7.90$\pm$0.05 & 0.60$\pm$0.02 & 0.010$\pm$0.0002 & 1.39$\pm$0.14 &  1.84$_{-0.05}^{+0.07}$&\\
WD8 & 24300$\pm$400 & 8.00$\pm$0.07 & 0.63$\pm$0.03 & 0.029$\pm$0.003  & 1.37$\pm$0.14 &  1.85$_{-0.05}^{+0.07}$&\\
WD9 & 20900$\pm$700 & 7.84$\pm$0.12 & 0.54$\pm$0.04 & 0.042$\pm$0.006  & 1.36$\pm$0.14 &  1.86$_{-0.05}^{+0.08}$&\\
\hline
\noalign{\smallskip}
NGC 6819 &  &  &  &  & & & $0.017$\\
\noalign{\smallskip}
WD6 & 21100$\pm$300 & 7.83$\pm$0.04 & 0.54$\pm$0.02 & 0.041$\pm$0.002 & 2.46$\pm$0.25 &  1.57$_{-0.05}^{+0.05}$&\\
WD7 & 16000$\pm$200 & 7.91$\pm$0.04 & 0.57$\pm$0.01 & 0.139$\pm$0.005 & 2.36$\pm$0.25 &  1.59$_{-0.05}^{+0.06}$&\\
\hline
\noalign{\smallskip}
Pleiades & & & & & & & $0.019$\\
\noalign{\smallskip}
WD0349$+$247 & 32841$\pm$172 & 8.63$\pm$0.04 & 1.03$\pm$0.02 & 0.048$\pm$0.004 & 0.071$\pm$0.008 &  5.87$_{-0.24}^{+0.31}$ &\\
\hline
\noalign{\smallskip}
Sirius B & 25000$\pm$200 & 8.60$\pm$0.04 & 1.0$\pm$0.01 & 0.108$\pm$0.003 & 0.129$\pm$0.012 & 4.67$_{-0.16}^{+0.18}$ & $0.020$\\
\hline
\noalign{\smallskip}
CPMPs  & & & & & & & \\
\noalign{\smallskip}
WD0315$-$011 & 7520$\pm$260  & 8.01$\pm$0.45 & 0.60$\pm$0.20 & 1.20$\pm$0.56   & 2.97$_{-2.12}^{+3.09}$ & 1.48$_{-0.28}^{+0.87}$ & 0.016$\pm$0.003 \\
WD0413$-$077 & 16570$\pm$350 & 7.86$\pm$0.05 & 0.54$\pm$0.02 & 0.112$\pm$0.008 & 0.96$\pm$0.37          & 2.07$_{-0.27}^{+0.53}$ & 0.008$\pm$0.001 \\
WD1354$+$340 & 13650$\pm$420 & 7.80$\pm$0.15 & 0.50$\pm$0.04 & 0.20$\pm$0.02   & 3.06$_{-1.46}^{+0.74}$ & 1.46$_{-0.09}^{+0.31}$ & 0.015$\pm$0.002 \\ 
WD1544$-$377 & 10600$\pm$250 & 8.29$\pm$0.05 & 0.78$\pm$0.02 & 0.76$\pm$0.05   & 0.18$\pm$0.50          & 4.13$_{-1.49}^{+?}$    & 0.021$\pm$0.003 \\
WD1620$-$391 & 24900$\pm$130 & 7.99$\pm$0.03 & 0.63$\pm$0.01 & 0.026$\pm$0.001 & 0.30$\pm$0.12          & 3.45$_{-0.35}^{+0.65}$ & 0.020$\pm$0.003 \\
WD1659$-$531 & 14510$\pm$250 & 8.08$\pm$0.03 & 0.66$\pm$0.01 & 0.24$\pm$0.01   & 2.27$_{-0.32}^{+0.34}$ & 1.58$_{-0.05}^{+0.08}$ & 0.019$\pm$0.004 \\
\hline
\hline
\end{tabular}}
\label{tab:mif2}
\end{center}
\end{table*}

\subsection{Open clusters and visual binaries}

We have carried out a re-analysis of the available data currently used to 
define the semi-empirical initial-final mass relationship, which is mainly 
based on white dwarfs in open clusters. We have used the white dwarf 
atmospheric parameters ($T_{\rm eff}$ and $\log g$) derived by other authors, 
as well as the ages and metallicities of the clusters reported in the 
literature. To obtain the final and initial masses we followed the procedure 
described in \citet{cat08}. This procedure consists in deriving the final mass 
($M_{\rm f}$) and the cooling time of each white dwarf from the atmospheric 
parameters and the cooling sequences of \citet{sal00}.  These cooling tracks 
consider a carbon-oxygen (CO) core white dwarf (with a larger abundance of O at 
the centre of the core) with a H thick envelope ontop of a He buffer, $q({\rm 
H})=M_{\rm H}/M=10^{-4}$ and $q({\rm He})=M_{\rm He}/M=10^{-2}$.  As it will be 
shown in the next section, the thicknesses of these envelops are very important 
in the cooling of white dwarfs.  These improved cooling sequences include an 
accurate treatment of the crystallization process of the CO core, including 
phase separation upon crystallization, together with up-to-date input physics 
suitable for computing white dwarf evolution.  Since we know the total ages of 
these white dwarfs (from the age of the cluster) we derived the main-sequence 
lifetimes of the progenitors, and from these, their initial masses using the 
stellar tracks of \citet{dom99}. At present, the atmospheric parameters of 
white dwarfs can be determined with accuracy if they correspond to the DA type, 
that is, if their spectra shows uniquely the hydrogen absorption lines.  For 
this reason, and in order to keep consistency in the cooling sequences used we 
only consider DA white dwarfs in our study. In Table \ref{tab:mif} we give 
the initial and final masses that we have recalculated for white dwarfs in open 
clusters. Other information such as the atmospheric parameters, cooling times, 
main-sequence lifetimes of the progenitors and metallicities are also given. 
Those white dwarfs for which we have obtained a cooling time longer than its 
total age have not been included in our study, since this is a good indication 
that the star does not belong to the cluster, or it has not a CO core. To 
compute the errors of the final masses we have taken into account the errors of 
the atmospheric parameters.  In the case of the initial masses we have taken 
into account the errors of the cooling times, which come from the atmospheric 
parameters, and the errors of the total ages of the white dwarfs. When no error 
for the total age was given in the literature, a value of 10 per cent was 
adopted. Finally, as reported in \citet{cat08}, to derive the initial masses of 
white dwarfs belonging to common proper motion pairs, the error of the 
metallicity was also taken into account.

\subsubsection*{NGC 2099 (M37)}

\citet{kal05} performed spectroscopic observations of 30 white dwarfs belonging 
to NGC 2099 (M37).  \citet{kal01} determined the age and metallicity of M37, 
$650$ Myr and $Z=0.011$, respectively. We assume an error of 10 per cent 
in the age of M37.

\subsubsection*{NGC 2168 (M35)}

We use the atmospheric parameters reported by Williams, Bolte \& Koester 
(2004). \citet{bar01} estimated the metallicity of this cluster, 
[Fe/H]$=-0.21\pm0.10$.  Thus, we use the stellar tracks corresponding to 
$Z=0.012$.  As in \citet{wil04} we use the age derived by \citet{von05}, 
$150\pm60$ Myr. The atmospheric parameters of LAWDS1 and LAWDS27 correspond to 
new data from \citet{fer05}.

\subsubsection*{NGC 3532}

The cluster data are from \citet{koe93}, although we have used the latest 
results of \citet{koe96} reported in \citet{fer05}. According to \citet{fer05} 
the age of this cluster is 300$\pm$150, and the metallicity is [Fe/H]$=-0.022$ 
(Twarog, Ashman \& Anthony-Twarog, 1997)  or [Fe/H]$=-0.02$ (Chen, Hou \& Hang, 
2003), so we have used the tracks corresponding to $Z=0.019$.  We have also 
included the data for three white dwarfs reported in \citet{rei89}, although 
the resolution is a bit lower.

\subsubsection*{Hyades and Praesepe}

We have used the data of \citet{cla01} for the Hyades and part of the Praesepe 
sample.  We have included also two stars studied by \citet{dob04} and some 
recent results on six new stars \citep{dob06}. According to \citet{von05}, the 
Hyades cluster has an age of 625$\pm$50 Myr. We assume the same value for 
Praesepe, since both belong to the same Hyades supercluster.  The metallicity 
is [Fe/H]$=0.13$ according to \citet{che03}, so we used the stellar tracks 
corresponding to $Z=0.027$.

\subsubsection*{NGC 2516}

We consider the atmospheric parameters derived by \citet{koe96}. The age of the 
cluster is that determined by Meynet, Mermilliod \& Maeder (1993), 141$\pm$2 
Myr. This cluster has a solar metallicity according to Jeffries, James \& 
Thurston (1998).

\subsubsection*{NGC 6791, NGC 7789 and NGC 6819}

We include the recent results obtained by \citet{kal08} based on spectroscopic 
observations of white dwarfs in different old clusters. These results, as well 
as the ones obtained by \citet{cat08} are somewhat relevant, since constitute 
the first constraints on the low-mass end of the initial-final mass 
relationship. NGC 6791 is one of the oldest and most metal-rich open clusters.  
According to \citet{kal07}, NGC 6791 has an age of 8.5 Gyr and 
[Fe/H]$=+0.3\pm0.5$. We have used the stellar tracks corresponding to $Z=0.04$. 
The white dwarfs of this cluster have relatively low masses 
($<0.47\,M_{\sun}$), so the majority of white dwarfs belonging to this cluster 
could have a He core \citep{kal07}.  Although it has been suggested that He-core white dwarfs could be formed also by single-evolution (Kilic, Stanek \& Pinsonneault, 2007), 
we prefer not to take these white dwarfs into account, and only consider the 
two confirmed cluster members with masses above this value. In this case, we assume an error of 10 per cent in the age 
of NGC 6791.

The white dwarf population of NGC 7789 and NGC 6819 were also studied recently 
by \citet{kal08}. According to their results, NGC 7789 has an age of 
1.4$\pm$0.14 Gyr and a metallicity of $Z=0.014$, while NGC 6819 has an age of 
2.5$\pm$0.25 Gyr and $Z=0.017$.

\subsubsection*{Pleiades}

WD0349$+$247 is the only known Pleiades white dwarf. We have used the 
atmospheric parameters derived by \citet{dob06}. The age of the Pleiades is 
125$\pm$8 Myr according to Rebolo, Martin \& Magazzu (1992) and it has a 
metallicity of [Fe/H]$=-0.03$ \citep{che03}. We use the stellar tracks 
corresponding to $Z=0.019$.

\subsubsection*{Sirius B}

Sirius B is a very well-known DA white dwarf that belongs to a visual binary 
system. Since the separation between the members of this system is $\sim7$ AU 
it can be assumed that there has not been any significant interaction between 
them during the AGB phase of the progenitor of the white dwarf member 
\citep{lie05a}. Thus, it can be considered that Sirius B has evolved as a 
single star, and therefore it is a good candidate for studying the 
initial-final mass relationship.  In this work, we use the atmospheric 
parameters determined by \citet{lie05a}, an age of 237$\pm$12 Myr and solar 
metallicity.

\subsection{Globular clusters}

In principle, globular clusters could also be used for improving the 
initial-final mass relationship since they have some characteristics which are 
similar to those of open clusters, but we have decided not to include them 
since the study of white dwarfs in globular clusters still suffers from large 
uncertainties.  Globular clusters contain thousands to millions of stars.  
However, white dwarfs belonging to them are usually very faint due to their 
large distances, which difficults obtaining high signal-to-noise spectra.  
Moreover, since globular clusters usually have crowded fields it is also 
difficult to isolate each star properly when performing spectroscopic 
observations. Therefore, the reduction procedure, mainly the background 
substraction, is also more complicated.  Up to now, the white dwarf population 
of globular clusters has been mainly used to determine their distances and ages 
\citep{ric04}.  An important attempt to derive the masses of white dwarfs in 
globular clusters (NGC 6397 and NGC 6752)  was carried out by \citet{moe04}. 
However, due to the low signal-to-noise and resolution of the spectra, they 
were not able to derive their spectroscopic masses independently and had to use 
a combination of spectra with photometry.  Although they calculated an average 
mass for the white dwarfs in NGC 6752 we have decided not to include this value 
in our work.  Instead, we prefer to use data of individual white dwarfs rather 
than a binned value for a cluster, contrary to that done in other works 
\citep{wil07,kal08}.

\subsection{Common proper motion pairs}

In the case of the common proper motion pairs, the procedure that we followed 
to derive the final and initial masses of the white dwarfs is explained in 
detail in \citet{cat08}.  It mainly consisted in performing independent 
spectroscopic observations of the components of several common proper motion 
pairs composed of a DA white dwarf and a FGK star. From the fit of the white 
dwarf spectra to synthetic models we derived their atmospheric parameters, and 
from these their masses and cooling times using the cooling sequences of 
\citet{sal00}. In order to derive the initial masses of the white dwarfs we 
performed independent high-resolution spectroscopic observations of their 
companions (FGK stars). Since it can be assumed that the members of a common 
proper motion pair were born simultaneously and with the same chemical 
composition \citep{weg73,osw88}, the ages and metallicities of both members of 
the pair should be the same.  From a detailed analysis of the spectra of the 
companions we derived their metallicites.  Then, we obtained their ages using 
either stellar isochrones, if the star was moderatly evolved, or the X-ray 
luminosity if the star was very close to the ZAMS (Ribas et al. in 
preparation). Once we had the total age of the white dwarfs and the metallicity 
of their progenitors we derived their initial masses using the stellar tracks 
of \citet{dom99}. In Table \ref{tab:mif2} we give the initial and final masses 
resulting from this study.

\section{The initial-final mass relationship}

\begin{figure}
\begin{center}
\includegraphics[scale=0.4]{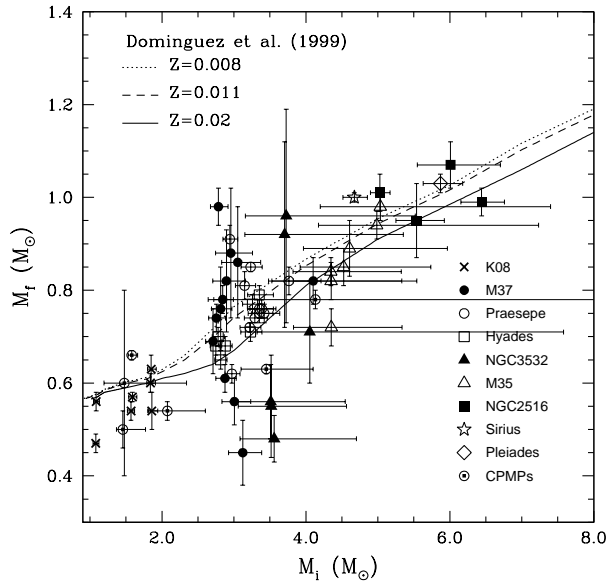}
\caption{Final masses  versus initial masses of  the available cluster
         and common proper motion pairs data.}
\label{fig:mif}
\end{center}
\end{figure}

In Fig.~\ref{fig:mif} we present the final versus the initial masses obtained 
for white dwarfs in common proper motion pairs and open clusters.  The 
observational data that can be used to define the semi-empirical initial-final 
mass relationship contains now 62 white dwarfs.  It is important to emphasize 
that all the values below $2.5~M_{\sun}$ correspond to our data obtained from 
common proper motion pairs (CPMPs) and the recent data obtained by \citet{kal08} --- K08.  
Before these studies no data for these small masses were available, since white 
dwarfs in stellar clusters are usually more massive, especially if the clusters 
are young.  The coverage of the low-mass end of the initial-final mass 
relationship is specially important since it guarantees, according to the 
theory of stellar evolution, the study of white dwarfs with masses near the 
typical values, $M\sim0.57\,M_{\sun}$, which represent about 90 per cent of the 
white dwarf population \citep{kep07}.  Thus, these new data increase the 
statistical significance of the semi-empirical initial-final mass relationship.

A first inspection of Fig.~\ref{fig:mif} reveals that there is a clear 
dependence of the white dwarf masses on the masses of their progenitors.  In 
Fig.~\ref{fig:mif}, we have also plotted the theoretical initial-final mass 
relationships of \citet{dom99} for different metallicities to be consistent 
with the stellar tracks used to derive the initial masses.  Although the 
distribution presents a large dispersion, a comparison of the observational 
data with these theoretical relationships shows that they share the same trend.  
However, it should be noted that for each cluster the data presents an 
intrinsic spread in mass.  The dispersion varies from cluster to cluster, but 
it is particularly noticeable for the case of M37.  Nevertheless, it should be 
taken into account as well that the observations of M37 were of poorer quality 
than the rest of the data \citep{fer05}.

\subsection{Main systematic uncertainties}

The results obtained are dependent on different assumptions and approaches that 
we have considered during the procedure followed to derive the final and 
initial masses, we discuss them separately.

\subsubsection{Thicknesses of the H and He envelopes}

The fact that the observed white dwarf masses in clusters scatter considerably 
in the same region of initial masses, as pointed out by other authors --- see, 
for instance, \citet{rei96} --- may indicate that mass loss could depend more 
on individual stellar properties than on a global mechanism, and that it could 
be a stochastic phenomenon, especially on the AGB phase.  Mass loss has a large 
impact on the final composition of the outer layers of white dwarfs, since it 
defines the thicknesses of the outer He and H (if present) layers. In fact, 
another reason that may explain why white dwarfs with different final masses 
could have progenitors with very similar initial masses is the assumption of a 
given internal composition and outer layer stratification of the white dwarfs 
under study.  The thicknesses of the H and He layers is a key factor in the 
evolution of white dwarfs, since they control the rate at which white dwarfs 
cool down. In this work we have used cooling sequences with fixed thicknesses 
of these envelopes, which might be more appropriate in some cases than in 
others.  In fact, the exact masses that the layers of H and He may have is 
currently a matter of debate being the subject of several studies.  For 
instance, \citet{pra02} computed models reducing the thickness of the H 
envelope to $q({\rm H})=2.32\times 10^{-6}$, obtaining cooling times shorter 
than those obtained in this work assuming a thick envelope ($q({\rm 
H})=10^{-4}$). This is natural since H has a larger opacity than He, and H is 
the major insulating component of the star. In the case of an even thinner H 
envelope ($q({\rm H})=10^{-10}$) the cooling age could be reduced in 10 per 
cent (1 Gyr) at $\log (L/L_{\sun})=-5.5$.  Thus, the uncertainty in the cooling 
times could be relevant in some cases, which would affect the estimates of the 
progenitor lifetimes and in turn, the initial masses derived.

In order to estimate the effect that the thicknesses of H and He envelopes may 
have in the initial masses derived here, we have repeated the calculations of 
section 2 but using the cooling sequences of Fontaine, Brassard \& Bergeron 
(2001) for a 50/50 CO core white dwarf with a standard He envelope, $q({\rm 
He})=10^{-2}$, and two different thicknesses for the H envelope, a thick one 
($q({\rm H})=10^{-4}$) and a thin one ($q({\rm H})=10^{-10}$). We have verified 
that the initial masses are indeed sensitive to the cooling sequences used, as 
expected. We have obtained larger initial masses when considering a thin 
envelope, due to the longer cooling times obtained in this case. As previously 
pointed out, in principle it can be expected that the cooling time scale should 
be smaller for a thin H envelope model, but this assumption is only true at low 
enough luminosities \citep{pra02}.  In fact, at intermediate luminosities a 
white dwarf with a thinner H envelope evolves slower than the thicker 
counterpart because it has an excess of energy to irradiate (Tassoul, Fontaine 
\& Winget, 1990). The maximum difference in the initial masses 
($\sim1~M_{\sun}$) has been found to occur for high-mass progenitors 
($M>5~M_{\sun}$), while this value is one order of magnitude smaller for 
smaller masses ($\sim0.1~M_{\sun}$). However, it should be noted that many 
other combinations are possible, for instance, with different thicknesses of 
the He envelope, which in this case has been kept fixed.  However, since it is 
impossible to know which is the real chemical stratification of the outer 
layers of each individual white dwarf we have not formally introduced this 
error in the calculations, since in some cases we would be overestimating the 
error of the initial masses derived in this work.

\subsubsection{Composition of the core}

It should be taken into account that besides a CO core, white dwarfs can have 
other internal compositions. Those white dwarfs more massive than 
$~1.05~M_{\sun}$ are thought to have a core made of ONe (Garc\'\i a-Berro, 
Ritossa \& Iben, 1997; Ritossa, Garc\'\i a-Berro \& Iben, 1996), while those 
with masses below $~0.4~M_{\sun}$ have an He-core.  ONe white dwarfs cool 
faster than CO or He white dwarfs because the heat capacity of O and Ne is 
smaller than that of C or He \citep{alt07}. On the contrary, He white dwarfs 
are the ones that cool slower \citep{ser02}.  Thus, those white dwarfs studied 
here with masses near the limits between different populations would be 
introducing an uncertainty in the cooling times obtained, since their cooling 
timescales are completely different from one composition to another. For 
example, a $~1~M_{\sun}$ ONe white dwarf cools 1.5 times faster than a CO white 
dwarf with the same mass \citep{alt07}. Thus, if an observed white dwarf has 
indeed a ONe core instead of the typical CO one, its progenitor lifetime would 
be underestimated in our analysis, and as a consequence, the initial mass 
derived would be more massive than the real one. However, the exact impact of 
this depends also on the total age of the white dwarf. The smaller the total 
age, the higher the effect of considering a wrong internal composition.

\subsubsection{Mass determinations when $T_{\rm eff}\leq12000$~K}

The errors reported in our study for the final masses only take into account 
the errors in the determination of the atmospheric parameters, which can be 
derived with accuracy if high signal-to-noise spectra are acquired.  However, 
it should be noted that this accuracy decreases considerably at low effective 
temperatures.  According to Bergeron, Wesemael \& Fontaine (1992), the 
atmospheres of DA stars below $12\,000$ K could be enriched in He while 
preserving their DA spectral type.  This He is thought to be brought to the 
surface as a consequence of the developement of a H convection zone. Depending 
on the efficiency of convection the star could still show Balmer lines, instead 
of being converted into a non-DA white dwarf.  Thus, the assumption of an 
unrealistic chemical composition could have a large impact on the cooling times 
estimated, but mainly at low effective temperatures. Nevertheless, a very large 
fraction of the stars in our sample ($\sim$95 per cent) have temperatures well 
above this limit.

\subsubsection{Total age of the white dwarfs}

As already pointed out, the derived initial masses depend on the cooling times, 
the total ages, the metallicity and finally, on the stellar tracks used.  
Among these parameters the largest source of error is due to the uncertainty in 
the total ages of the white dwarfs. For white dwarfs in open clusters, the age 
can be usually derived with high accuracy from model fits to the turn-off 
location in a colour-magnitude diagram.  The uncertainty on the age of a 
cluster is a systematic effect for stars belonging to the same cluster, since 
all the initial masses will be shifted together to larger or smaller masses in 
the final versus initial masses diagram \citep{wil07}. On the contrary, in the 
case of white dwarfs in common proper motion pairs, the accuracy in the total 
age depends on the evolutionary stage of the companion.  The accuracy of the 
age using isochrone fitting could be high if the star is relatively evolved and 
located far away from the ZAMS.  Using the X-ray luminosity method, described 
in \citet{cat08}, the ages derived could also be quite precise (from 8 to 20 
per cent) if the star is relatively young ($t\leq1$~Gyr).

\subsection{The semi-empirical relationship}

Following closely recent works on this subject \citep{fer05,wil07,kal08}, we 
assume that the initial-final mass relationship can be described as a linear 
function. We have performed a weighted least-squares linear fit of the data, 
obtaining that the best solution is

\begin{equation}
M_{\rm f}=(0.117\pm0.004)M_{\rm i}+(0.384\pm0.011)
\label{eq:mf_mi1}
\end{equation}

\noindent where the errors are the standard deviation of the coefficients.

\begin{figure}
\begin{center}
\includegraphics[scale=0.4]{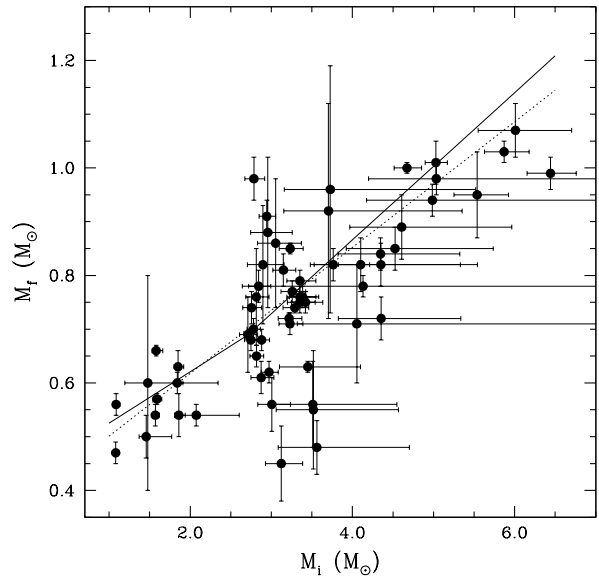}
\caption{Final masses  versus initial masses  for the white  dwarfs in
         our  sample. Solid  and dotted  lines correspond  to weighted
         least-squares linear fits of the data.}
\label{fig:mif2}
\end{center}
\end{figure}

In Fig.~\ref{fig:mif2} we represent all the data that we have re-calculated and 
this linear fit (dotted line). In past works, since there was not available 
data in the region of low-mass white dwarfs, a least-squares linear fit led to 
an unconstrained result \citep{fer05}. For this reason, a ficticious anchor 
point at low masses was used to represent the typical white dwarf mass of 
$M_{\rm f}\sim0.57~M_{\sun}$ \citep{kep07}. In our case, this is not necessary 
since we are now reproducing this well-established peak of the field white 
dwarf mass distribution thanks to the new data in the low-mass region 
\citep{kal08,cat08}.  As can be seen in Fig.~\ref{fig:mif2} the theoretical 
initial-final mass relationship can be divided in two different linear 
functions, each one above and below 2.7~$M_{\sun}$, with a shallower slope for 
small masses probably due to the smaller efficiency of mass loss.  Taking this 
into account we have performed a weighted least-squares linear fit for each 
region, obtaining

\begin{equation}
M_{\rm f} = (0.096\pm0.005)M_{\rm i}+(0.429\pm0.015) \\
\label{eq:mf_mi2}
\end{equation}

\noindent   for   $M_{\rm   i}<2.7~M_{\sun}$,  whereas   for   $M_{\rm
i}>2.7~M_{\sun}$ we obtain:

\begin{equation}
M_{\rm f} = (0.137\pm0.007)M_{\rm i}+(0.318\pm0.018)
\label{eq:mf_mi3}
\end{equation}

In these expressions the errors are the standard deviation of the coefficients.  
These two independent fits, which are represented as solid lines in 
Fig.~\ref{fig:mif2}, seem to reproduce better the observational data than a 
unique linear fit (dotted line).  Taking into account the scatter of the data 
and the values of the reduced $\chi^2$ of these fits (7.1 and 4.4, 
respectively) we consider that the errors associated to the coefficients are 
underestimated. A more realistic error can be obtained computing the dispersion 
of the derived final masses, which is of $0.05~M_{\sun}$ and $0.12~M_{\sun}$ 
respectively.  These are the errors that should be associated to the final mass 
when using the expressions derived here --- Eqs.~(\ref{eq:mf_mi2}) and 
(\ref{eq:mf_mi3}), respectively.

\subsection{Dependence on different parameters}

As already mentioned in the introduction, there are other parameters besides 
the mass of the progenitor that may have an impact on the final masses of white 
dwarfs (e.g.~metallicity or rotation).  A detailed analysis of the results that 
we have obtained so far allows us to give some clues on the dependence of the 
initial-final mass relationship on these parameters. We discuss them below.

\subsubsection{Metallicity}

The sample of white dwarfs studied here covers a range of metallicities from 
$Z=0.006$ to $0.040$.  From a theoretical point of view it is well established 
that progenitors with large metallicity produce less massive white dwarfs --- 
see the relations of \citet{dom99} plotted in Fig.~\ref{fig:mif}.  Thus, one 
should expect to see a dependence of the semi-empirical data on metallicity. 
Our purpose in this section is to compare data with the same and different 
metallicity and evaluate if the differences in the derived masses are smaller 
or greater in this two cases. Open clusters are appropriate for carrying out 
such comparison, since all the stars belonging to a particular cluster have the 
same metallicity. For instance, in the case of the two stars from Praesepe 
(open circles in Fig.  \ref{fig:mif}) with initial masses around 
$3.0~M_{\sun}$, the white dwarfs differ in $\Delta M_{\rm f}=0.3~M_{\sun}$.  
However, for the rest of stars in this cluster, the dispersion is significantly 
smaller (a factor of 2).  Thus, the large spread could be explained by 
considering that may be these objects are field stars.  However, the sample of 
white dwarfs in the Praesepe cluster has been well studied \citep{cla01,dob04} 
and it is unlikely that these stars do not belong to the cluster. The Hyades 
also contain a large number of white dwarfs, for which the initial and final 
masses have been derived with accuracy. In this case, all the data points fall 
in the region limited by the theoretical relations of \citet{dom99}, and the 
scatter is $\Delta M_{\rm f}=0.1~M_{\sun}$, which is the same as the difference 
between the theoretical relations corresponding to $Z=0.008$ and $Z=0.02$. This 
is the minimum scatter found in the observational data, since the points 
corresponding to the Hyades, together with those of Praesepe, are the ones with 
smaller error bars.  In any case, this scatter, although smaller than in other 
cases, is still larger than the uncertainties, which prevents to derive any 
clear dependence on metallicity.  As previously pointed out, the data of M37 
present the largest scatter in the sample ($\Delta M_{\rm f}=0.5~M_{\sun}$), 
but it should be noticed that the errors in the final masses are considerably 
larger than for the rest of the clusters.

It is important to evaluate if the dispersion increases when data with 
different metallicities is considered.  Comparing points with the same initial 
mass but different metallicities, it can be noted that $\Delta M_{\rm f}$ is of 
the same order as in the case of equal metallicities. For example, comparing 
the data of M37 and the data of the Hyades --- which have metallicities 
$Z=0.011$ and $Z=0.027$, respectively --- for an initial mass around 
$3~M_{\sun}$, $\Delta M_{\rm f}$ is $\sim 0.5~M_{\sun}$, but this scatter is 
the same when we compare data from M37 only.  However, the scatter decreases 
when we analyse larger initial masses.  If the data of M35 and M37 are compared 
(metallicites $Z=0.011$) with the data of NGC3532 and one of the common proper 
motion pairs located at $\sim 4.0~M_{\sun}$ (with $Z=0.02$), it can be seen 
that the maximum $\Delta M_{\rm f}$ is $0.2~M_{\sun}$, although this is also 
what we obtain when we compare only the data of M35 located in this region.  
In the region of small masses we have data ranging from $Z=0.008$ to $Z=0.04$, 
but the maximum scatter is the same regardless of the metallicity of the stars 
($\Delta M_{\rm f}=0.1~M_{\sun}$). Finally, at the high-mass end we find the 
same dispersion, although in this case the data have the same metallicity.  
Thus, considering the current accuracy of the available observational data we 
do not find any clear dependence of the semi-empirical initial-final mass 
relationship on metallicity.
	
\begin{figure}
\begin{center}
\includegraphics[scale=0.4]{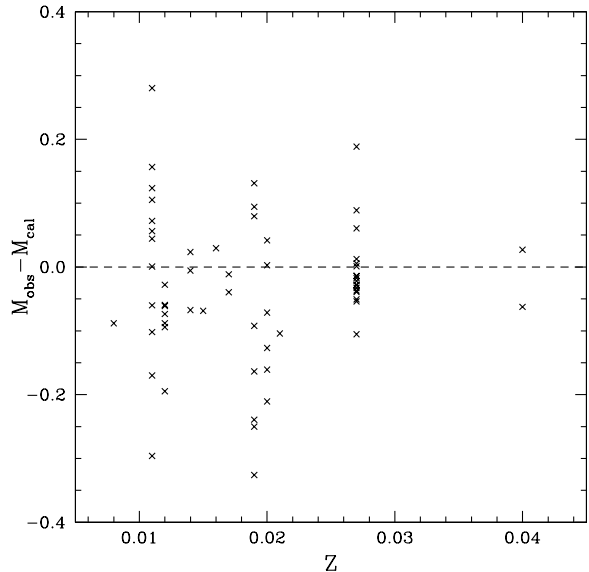}      
\caption{Correlation between final masses and metallicity.}
\label{fig:mf_cor}
\end{center}
\end{figure}

It is interesting to perform a more quantitative study of the correlation 
between the final masses and metallicity.  In Fig.~\ref{fig:mf_cor} we plot the 
differences between the observed final masses and the final masses obtained 
using Eqs.~(\ref{eq:mf_mi2})  and ~(\ref{eq:mf_mi3})  as a function of 
metallicity.  The dashed line corresponds to the hypothetical case in which 
there is no difference between the observational and the values predicted by 
these relations.  In order to quantify the correlation in the sample of points 
presented in Fig.~\ref{fig:mf_cor}, we have calculated the Spearman rank 
correlation coefficient.  The value obtained is 0.036, which is very close to 
zero, indicating that there is an extremely weak positive correlation between 
the difference in the final masses and metallicity.  Since the Spearman 
correlation does not take into account the errors of the values considered, we 
have carried out a bootstrapping in order to evaluate the actual uncertainty on 
the correlation coefficient without any assumption on the error bars.  This 
consists on choosing at random $N$ objects from our sample (which has also $N$ 
objects) allowing repetition, and then calculating the new correlation 
coefficient for each of these new samples.  We have performed this a large 
number of times (5000) obtaining a mean correlation coefficient of $-0.002\pm 
0.128$.  Thus, we conclude that the final masses and metallicity of this sample 
does not present any correlation, and that the scatter in the distribution in 
Fig.~\ref{fig:mif} is not due to the effect of metallicity. Of course, it 
should be taken into account that the observed final masses have been derived 
using the atmospheric parameters reported by different authors who have 
considered also different white dwarf models, although the prescriptions used 
in the fits are usually those of \citet{ber92} in all the cases.

It is worth comparing our results with other works.  For instance, 
\citet{kal05} claimed that they had found the first evidence of a metallicity 
dependence on the initial-final mass relationship. They noticed that half of 
their data of M37 (also plotted in Fig.~\ref{fig:mif})  were in agreement with 
the theoretical relationship of \citet{mar01}, and considered this result as an 
indication of dependence on metallicity.  On the other hand, in a recent 
revision of the semi-empirical initial-final mass relationship, \citet{wil07} 
analyzed part of the cluster data discussed here by deriving a binned 
semi-empirical initial-final mass relationship. This consisted in associating 
an initial and a final mass for each cluster by calculating the mean of the 
initial and final masses of the individual white dwarfs belonging to that 
cluster. Then, they compared these values as function of metallicity and, as in 
our case, they did not find a clear dependence of the semi-empirical data on 
metallicity. In fact, their conclusion was that metallicity should affect the 
final mass only in $0.05~M_{\sun}$, considering an initial mass of 
$3~M_{\sun}$, which is in good accord with our results.

\subsubsection{Rotation}

According to \citet{dom96}, fast rotating stars produce more massive white 
dwarfs than slow rotating stars.  The models calculated by \citet{dom96} 
including rotation predict that for a fast rotating star with an initial mass 
of $6.5~M_{\sun}$ the white dwarf produced has a mass in the range $1.1$ to 
$1.4~M_{\sun}$, which is considerably larger than when rotation is disregarded 
\citep{dom99}.  Among our sample of white dwarfs in common proper motion pairs 
there are two stars (WD1659$-$531 and WD1620$-$391)  that might exemplify the 
effect that rotation may have in stellar evolution.  The companion of 
WD1659$-$531, HD153580, is a fast rotating star according to \citet{rei03}, 
with a tangential velocity $v\sin i= 46\pm5 $ km s$^{-1}$.  Thus, we can 
hypotetically assume that the progenitor of WD1659$-$531 was also a fast 
rotator. If we compare the masses derived for these stars (Table 
\ref{tab:mif2}) we can note that starting from an initial mass of 
$1.58~M_{\sun}$, which is more than two times smaller than the progenitor of 
WD1620$-$391 ($3.45~M_{\sun}$), it ends up as a white dwarf with approximately 
the same mass, $0.66~M_{\sun}$.  This indicates that the progenitor of 
WD1659$-$531 lost less mass during the AGB phase than the progenitor of 
WD1620$-$391. These differences may not be related to metallicity, since both 
progenitors had solar composition. Thus, we think that this could be the first 
evidence showing that rotation may have a strong impact in the evolution of a 
star, leading to more massive white dwarfs, as suggested by \citet{dom96}.

\subsubsection{Magnetism}

One way to detect the presence of magnetic fields in white dwarfs is by 
performing spectropolarimetric observations.  Some of the white dwarfs 
belonging to this sample have been the subject of studies to investigate their 
magnetic nature.  In particular, WD 0837$+$199 (also known as EG61), which 
belongs to Praesepe, is the only known magnetic white dwarf in an open cluster. 
The magnetic field of this star is of 3 MG according to the study of 
\citet{kaw07}. The mass that we have obtained for this star is 
$0.81~M_{\sun}\pm0.03~M_{\sun}$ (see Table \ref{tab:mif2}), rather large, 
although a bit smaller than the typical mass of magnetic white dwarfs, which is 
around $0.93~M_{\sun}$ \citep{wic05}.  Regarding white dwarfs in common proper 
motion pairs, \citet{kaw07} also obtained circularly polarized spectra of 
WD0413$-$017, WD1544$-$377, WD1620$-$391 and WD1659$-$531, finding evidence of 
magnetism only for the first of these stars.  This result is in good agreement 
with the findings of other authors \citep{azn04,jor07}. In the case of 
WD0413$-$017, more commonly known as 40 Eri B, the magnetic field is rather 
weak, 2.3 kG, and the mass that we have derived is $0.54\pm0.02~M_{\sun}$. 
Although the mass of this star is well below the typical mass of magnetic white 
dwarfs, it is in good agreement with the rest of white dwarfs studied by 
\citet{kaw07} at the kG level. If we do not consider WD 0837$+$199 and 
WD0413$-$017 in the fit carried out in the last section we obtain negligible 
changes on the expressions derived.

Spectropolarimetric surveys of white dwarfs have suggested that there is a 
decline in the incidence of magnetism of stars with fields $B<10^6$ G, although 
this incidence seems to rise again when the field is much lower, $B<100$ kG 
\citep{wic05}. The sample of white dwarfs that we have considered in this work 
contains two magnetic white dwarfs, one with a strong magnetic field and the 
other with a rather weak magnetic field.  Although the influence of the 
magnetic field in stellar evolution has not been yet established, we find that 
the final masses obtained in these two cases are very different, being larger 
when the magnetic field is stronger. So, it is reasonable to think that the 
magnetic field could play a key role on the evolution of the progenitor star. 
However, with the current data on the magnetic white dwarfs belonging to the 
sample of white dwarfs used in this work it is not possible to favor one of the 
two main hypothesis regarding this issue:  whether magnetic white dwarfs are 
more massive because the progenitors were also more massive (without any 
dependence on the magnetic field), or on the contrary, magnetic white dwarfs 
are more massive because the magnetic field had an influence during its 
evolution, favoring the growth of the core.

According to \citet{kaw03}, 16 per cent of the white dwarf population should be 
comprised by magnetic white dwarfs. So, among the sample of stars considered in 
this work, we could expect 9-10 white dwarfs to be magnetic. Thus, 
spectropolarimetric observations of the current sample of white dwarfs used to 
define the semi-empirical initial-final mass relationship would be very useful 
to shed some light upon this subject.

\section{The white dwarf luminosity function}

The white dwarf luminosity function is defined as the number of white dwarfs 
per unit volume and per bolometric magnitude --- see, for instance, 
\citet{ise98}:

\begin{equation}
n(M_{\rm bol},T)=\int_{M_{\rm i}}^{M_{\rm s}}
\phi(M) \psi(T-t_{\rm cool}-t_{\rm prog})\tau_{\rm cool}\;dM
\label{eq:wdlf}
\end{equation}

\noindent where $M$ is the mass of the progenitor of the white dwarf, 
$\tau_{\rm cool}=dt/dM_{\rm bol}$ is its characteristic cooling time, $M_{\rm 
i}$ and $M_{\rm s}$ are the minimum and maximum mass of the progenitor star 
able to produce a white dwarf with a bolometric magnitude $M_{\rm bol}$ at time 
$T$, $t_{\rm cool}$ is the time necessary to cool down to bolometric magnitude 
$M_{\rm bol}$ --- for which we adopt the results of \citet{sal00} and 
\citet{alt07} for CO and ONe white dwarfs, respectively --- $t_{\rm prog}$ is 
the lifetime of the progenitor, $T$ is the age of the population under study, 
$\psi(t)$ is the star formation rate --- which we assume to be constant --- and 
$ \phi(M)$ is the initial mass function, for which we adopt the expression of 
\citet{sal55}.

\subsection{The influence of the progenitors}

\begin{figure}
\begin{center}
\includegraphics[scale=0.4]{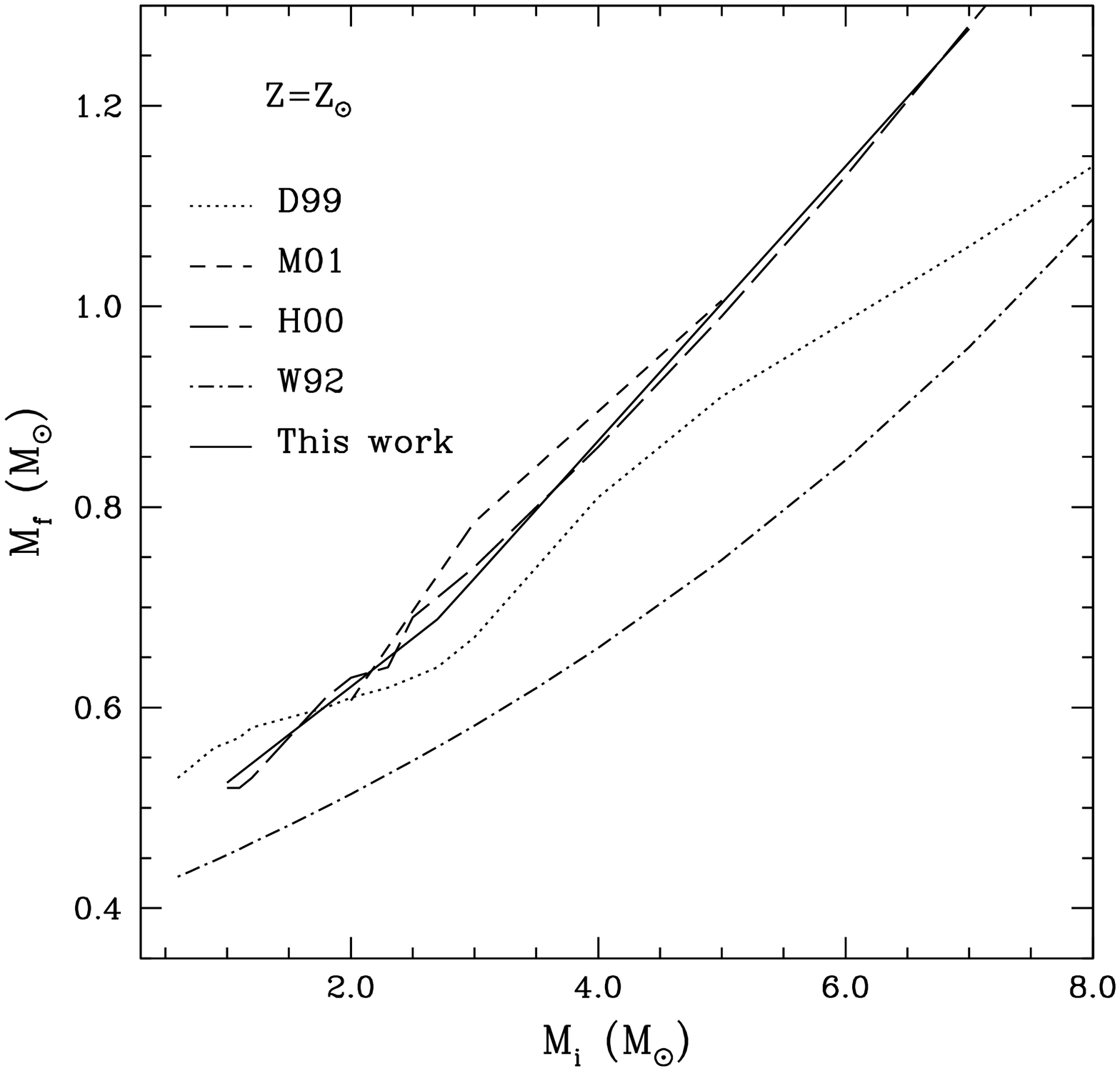}
\caption{Initial-final   mass  relationship  according   to  different
        authors: \citet{dom99}  --- D99 --- \citet{mar01}  --- M01 ---
        \citet{hur00} ---  H00 --- \citet{woo92}  --- W92 ---  and the
        one derived in this work.}
\label{fig:ifmr}
\end{center}
\end{figure}

\begin{figure}
\begin{center}
\includegraphics[scale=0.4]{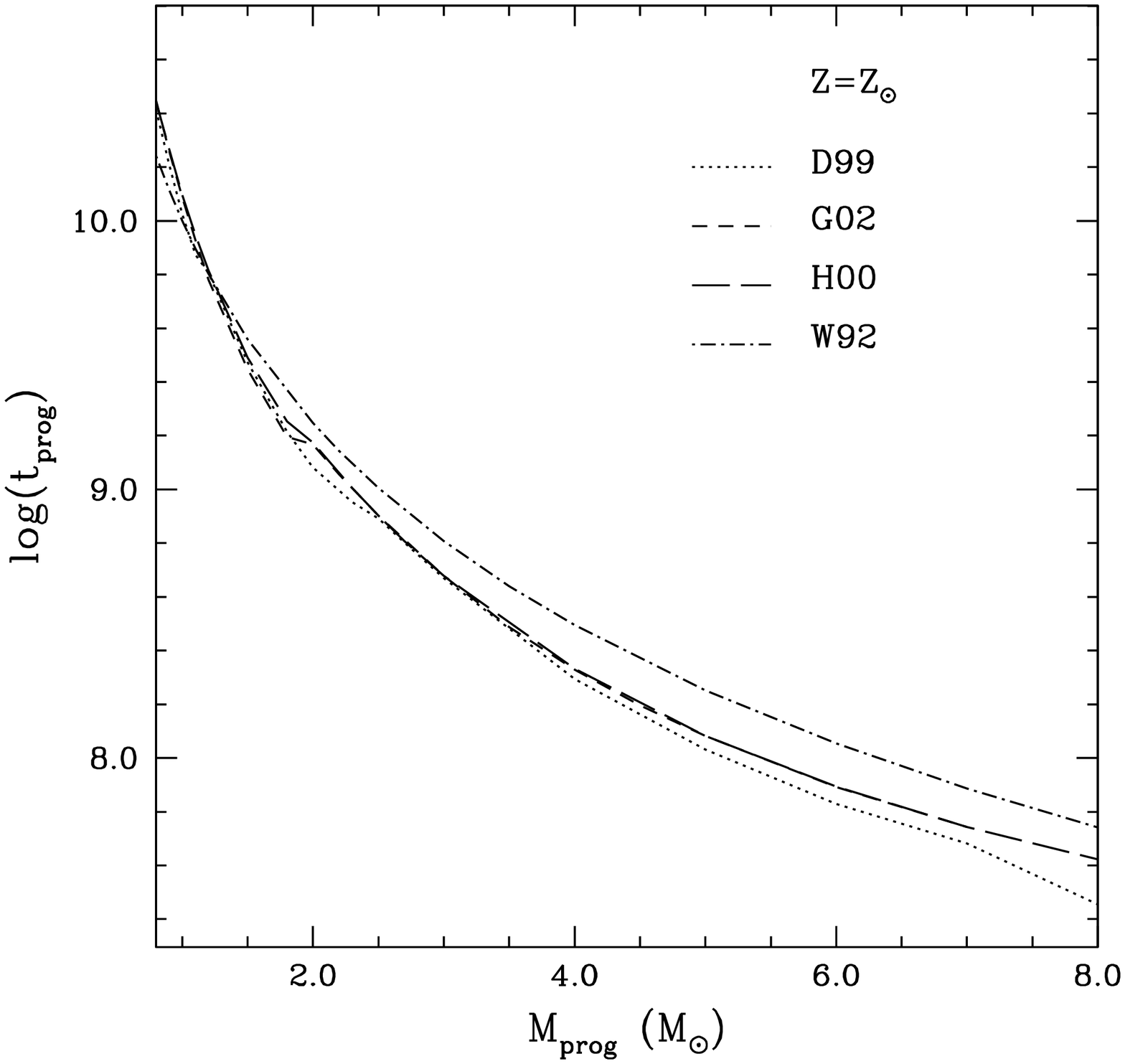}
\caption{Main  sequence  lifetime  versus  stellar mass  according  to
         different  authors: \citet{dom99}  --- D99  --- \citet{gir02}
         --- G02 ---  \citet{hur00} --- H00 ---  and \citet{woo92} ---
         W92.}
\label{fig:mslife}
\end{center}
\end{figure}

To compute the white dwarf luminosity function it is also necessary to provide 
a relationship between the mass of the progenitor and the mass of the resulting 
white dwarf, that is, the initial-final mass relationship.  Additionally, the 
influence of the progenitors in Eq.~(\ref{eq:wdlf})  appears through the age 
assigned to the progenitor, which determines the star formation rate, and 
through the cooling time, $t_{\rm cool}$ and the characteristic cooling time, 
$\tau_{\rm cool}$, which depend on the mass of the white dwarf. In order to 
evaluate the influence of these inputs, we have computed a series of 
theoretical white dwarf luminosity functions using several initial-final mass 
relationships (Fig.~\ref{fig:ifmr})  and evolutive tracks for the progenitor 
stars (Fig.~\ref{fig:mslife}). Whenever possible we have adopted the same set 
of stellar evolutionary inputs, that is, the initial-final mass relationship 
and the main-sequence lifetime corresponding to the same set of calculations.

We compare the resulting theoretical luminosity functions with the data 
obtained by averaging the different observational determinations of the white 
dwarf luminosity function (Knox, Hawkins \& Hambly, 1999; Leggett, Ruiz \& 
Bergeron, 1998; Oswalt et al. 1996; Liebert, Dahn \& Monet, 1988). The 
theoretical white dwarf luminosity functions were also normalized to the 
observational value with the smallest error bars in number density of white 
dwarfs, that is $\log(N)=-3.610$, $\log(L/L_{\sun})=-2.759$, avoiding in this 
way the region in which the cooling is dominated by neutrinos (at large 
luminosities)  and the region in which crystallization is the dominant physical 
process --- at luminosities between $\log(L/L_{\sun})\simeq -3$ and $-4$.

\begin{figure}
\begin{center}
\includegraphics[scale=0.4]{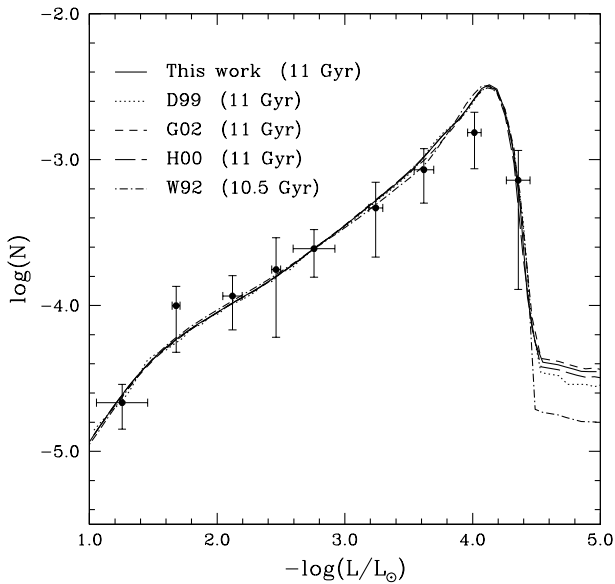}
\caption{White   dwarf  luminosity  functions   considering  different
         evolutive    stellar    models    and   initial-final    mass
         relationships: \citet{dom99} --- D99--- \citet{gir02} --- G02
         --- \citet{hur00} ---  H00 --- \citet{woo92} ---  W92 --- and
         the relation derived in this work. See text for details.}
\label{fig:Lum-var}
\end{center}
\end{figure}

Fig.~\ref{fig:Lum-var} shows the resulting white dwarf luminosity functions 
when different stellar evolutionary inputs are used. At low luminosities it can 
be noticed the characteristic sharp down-turn in the density of white dwarfs.  
This cut-off in the number counts has been interpreted by different authors 
\citep{win87,gar88} as the consequence of the finite age of the Galactic disc. 
Thus, a comparison between the theoretical luminosity functions and the 
observational data can provide information about the age of the Galactic disc.  
In this figure the cut-off of the observational white dwarf luminosity function 
has been fitted using an age of the disc of $T=11$ Gyr for all the cases except 
for the case in which the expressions of \citet{woo92} were used. In this last 
case the best-fitting is obtained using $T=10.5$ Gyr.  This can be understood 
by comparing the different stellar evolutionary inputs considered in this work.  
As can be seen in Fig.~\ref{fig:ifmr}, the initial-final mass relationship of 
\citet{woo92} is the one that produces less massive white dwarfs. The 
semi-empirical relationship that we have derived in this work is similar to 
that of Hurley, Pols \& Tout (2000).  \citet{mar01} predicts more massive white 
dwarfs at the intermediate mass domain, whereas the results of \citet{dom99} 
produce more massive remnants for the low-mass end and less massive white 
dwarfs for the high mass end, but always substantially larger than those 
obtained with the initial-final mass relationship of \citet{woo92}.  These 
differences mainly arise from the procedure used to calculate each theoretical 
initial-final mass relationship.  The relations derived by \citet{dom99} and 
\citet{mar01} are obtained using fully evolutive models but using different 
treatments of convective boundaries, mixing and mass-loss rates on the AGB 
phase.  On the other hand, the relations of \citet{hur00} were obtained from a 
fitting of the observational data from eclipsing binaries and open clusters, 
obtaining analytic formulae for different metallicities.  Finally, the relation 
of \citet{woo92} is an exponential expression derived from a fit to the PNN 
mass distribution.  In Fig.~\ref{fig:mslife} we show the different 
main-sequence lifetimes as a function of the main-sequence mass.  In those 
cases in which there was a dependence on metallicity we have adopted 
$Z=Z_{\sun}$.  As it can be seen there, the behaviour of the different 
main-sequence lifetimes is very similar, although for the case of 
\citet{woo92}, stars spend more time in the main-sequence than in the rest of 
cases, especially for those stars with large masses. Considering this, one 
should expect that the fit of the cut-off of the white dwarf luminosity 
function would correspond to a longer Galactic disc age when using the 
expressions of \citet{woo92}.  On the contrary, we have obtained a younger 
Galactic disc. The longer progenitors lifetimes of \citet{woo92} are in part 
compensated by the fact that its corresponding initial-final mass relationship 
favors the production of low-mass white dwarfs, which cool down faster at high 
luminosities.  A simple test can be done by using, for example, the stellar 
tracks of \citet{dom99}, which give shorter progenitor lifetimes, and the 
initial-final mass relationship of \citet{woo92}.  In this case we obtain a 
Galactic disc age of 10 Gyr, 1 Gyr younger than if we consider any of the other 
initial-final mass relationships shown in Fig.~\ref{fig:ifmr}.  Thus, the 
behaviour of the relation of \citet{woo92} is clearly different from others in 
the literature, and this has important implications on the resulting white 
dwarf luminosity functions.

Comparing the results of Fig.~\ref{fig:Lum-var} it can be noticed that for the 
hot end of the white dwarf luminosity function there are not differences 
whatsoever.  In fact, all the theoretical luminosity functions are remarkably 
coincident.  The only visible differences, although not very relevant, occur 
just after the crystallization phase has started, at $\log(L/L_{\sun})\simeq 
-4.0$. Two essential physical processes are associated with crystallization, 
namely, a release of latent heat and a modification of the chemical 
concentrations in the solid phase.  Both provide extra energy sources and 
lengthen the cooling time of the star. This is the reason why all the 
theoretical calculations predict a larger number of white dwarfs for these 
luminosity bins. Beyond the cut-off, we find that the density of white dwarfs 
is smaller in the case of \citet{woo92}, and this is because this region is 
dominated by massive white dwarfs, since the progenitors of these stars were 
also massive, and as shown in Fig.~\ref{fig:mslife} spent less time at the 
main-sequence.  Thus, massive white dwarfs have had enough time to cool down to 
such low luminosities. In any case, it is worth mentioning that low-mass white 
dwarfs cool faster at high luminosities, but for small luminosities it occurs 
just the opposite, since massive white dwarfs crystallize at larger 
luminosities, and this implies smaller time delays.

\subsection{The luminosity function of massive white dwarfs}

\subsubsection{Effect of the initial-final mass relationship}

\begin{figure}
\begin{center}
\includegraphics[scale=0.4]{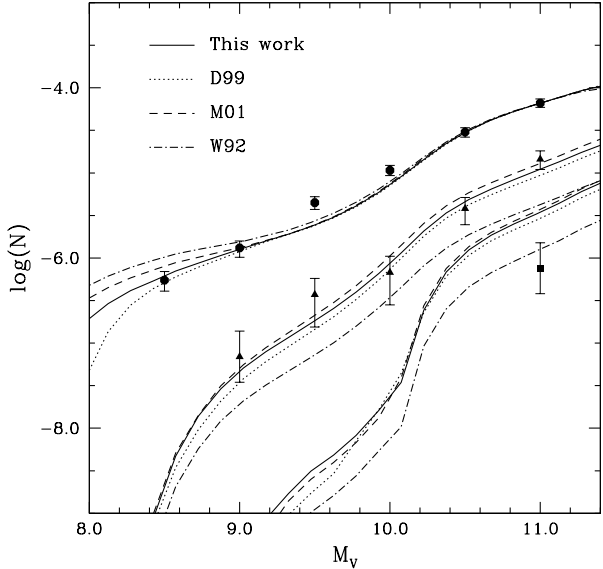}
\caption{White  dwarf  luminosity  functions versus  visual  magnitude
 	 using    different    initial-final    mass    relationships:
 	 \citet{dom99}   ---  D99  ---   \citet{mar01}  ---   M01  ---
 	 \citet{woo92} --- W92  --- and this work. From  top to bottom
 	 we  show the  total luminosity  function, and  the luminosity
 	 functions   of   white  dwarfs   with   masses  larger   than
 	 $0.7~M_{\sun}$  and $1.0~M_{\sun}$.   Circles,  triangles and
 	 squares   correspond    to   the   observational    data   of
 	 \citet{lie05b}.}
\label{fig:Lum-mag}
\end{center}
\end{figure}

The influence of the initial-final mass relationship on the white dwarf 
luminosity function should be more evident when it is constrained to massive 
white dwarfs \citep{dia94}.  Recently, Liebert, Bergeron \& Holberg (2005b) 
performed high signal-to-noise spectroscopic observations of more than 300 
white dwarfs belonging to the Palomar Green (PG) Survey.  The analysis of this 
set of data has provided us with a sample of white dwarfs with well determined 
masses that allows for the first time the study of the white dwarf luminosity 
function of massive white dwarfs \citep{ise07}.  Unfortunately, according to 
\citet{lie05b} the completeness of the sample decreases severely near 
$10\,000$~K, where white dwarfs with small masses ($0.4~M_{\sun}$) are brighter 
($M_{\rm V}\sim11$)  than massive white dwarfs with $M>0.8~M_{\sun}$ (which 
have visual magnitudes around $M_{\rm V}\sim13$).  Consequently, above $M_{\rm 
V}=11$ this survey only has detected white dwarfs with masses larger than 
$0.4~M_{\sun}$. Therefore, we will limit the analysis to white dwarfs brighter 
than $M_{\rm V} \sim 11$.

\begin{figure}
\begin{center}
\includegraphics[scale=0.4]{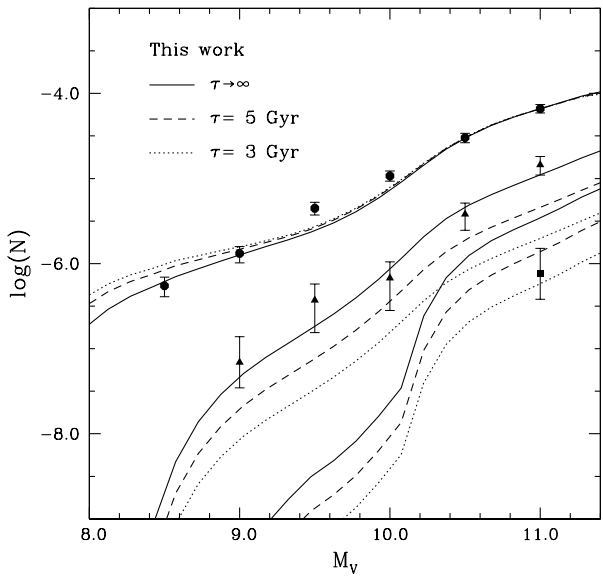}
\caption{White  dwarf  luminosity  functions versus  visual  magnitude
         using  the initial-final  mass relationship  derived  in this
         work and  different star formation  rates. Circles, triangles
         and   squares  correspond  to   the  observational   data  of
        \citet{lie05b}.}
\label{fig:Lum-sfrmag1}
\end{center}
\end{figure}

\begin{figure}
\begin{center}
\includegraphics[scale=0.4]{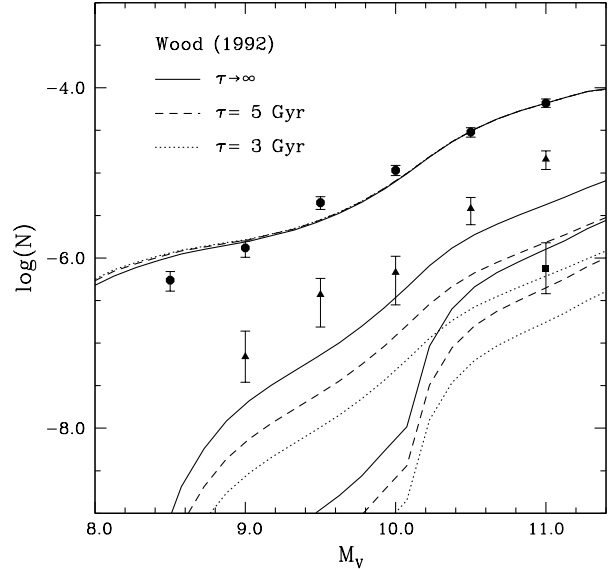}
\caption{Same    as   Fig.~\ref{fig:Lum-sfrmag1}    but    using   the
         initial-final mass relationship of Wood (1992).}
\label{fig:Lum-sfrmag2}
\end{center}
\end{figure}

We have computed a set of white dwarf luminosity functions considering an age 
of 11 Gyr for the Galactic disc and using bins of visual magnitude. In 
Fig.~\ref{fig:Lum-mag} we show from top to bottom the total luminosity function 
and the luminosity functions of white dwarfs with masses larger than 
$0.7~M_{\sun}$ and $1.0~M_{\sun}$, respectively.  The total luminosity function 
(that is, considering the whole the range of masses)  was normalized to the bin 
corresponding to $M_{\rm V}=11$, and then, this normalization factor was used 
for the luminosity functions of white dwarfs more massive than $0.7~M_{\sun}$ 
and $1.0~M_{\sun}$.  In this case, we have used the stellar evolutionary inputs 
of \citet{dom99}, \citet{mar01}, \citet{woo92} and the semi-empirical 
initial-final mass relationship that we have derived in the previous section. 
Comparing the different theoretical luminosity functions, it can be noted that 
the predicted number of massive white dwarfs is larger when using the inputs of 
\citet{dom99}, \citet{mar01} and our semi-empirical initial-final mass 
relationship in comparison with the results obtained when considering the 
expressions of \citet{woo92}.  This is obviously due to the fact that the 
initial-final mass relationship of \citet{woo92} favors the production of 
low-mass white dwarfs.  It can also be noted that the density of massive white 
dwarfs is slightly larger when considering the initial-final mass relationship 
of \citet{mar01} than that obtained when using the initial-final mass 
relationship derived here. The reverse is true when the initial-final mass 
relationship of \citet{dom99} is used. Without considering the results obtained 
when the expressions of \citet{woo92} are used, it can be noted that it is not 
possible to evaluate which initial-final mass relationship produces a 
theoretical luminosity function that better fits the observational data, since 
the error bars of the observational data are larger than the differences 
between the theoretical results.  In any case, what it can be clearly seen is 
that all the theoretical relations predict more massive white dwarfs than the 
observations when a mass cut of $1.0~M_{\sun}$ is adopted, except in the case 
of \citet{woo92}.

\subsubsection{Effect of the star formation rate}

The star formation rate considered in our calculations has also an important 
influence on the number of massive white dwarfs produced. In our previous 
calculations we have considered a constant star formation rate, which has led 
us to obtain more massive white dwarfs than those detected in the PG Survey.  
To evaluate the effect of the star formation rate on the white dwarf luminosity 
function we have repeated the calculations considering the semi-empirical 
initial-final mass relationship derived in this work and an exponentially 
decreasing star formation rate $\psi(t)=\exp{(-t/\tau)}$, with $\tau=3$ and $5$ 
Gyr, respectively.  As it can be noted from Fig.~\ref{fig:Lum-sfrmag1} the 
production of massive stars decreases considerably when a exponentially 
decreasing star formation rate is considered.  The observational data 
corresponding to the whole range of white dwarf masses and to masses larger 
than $0.7~M_{\sun}$ is better fitted when a constant star formation rate 
($\tau\rightarrow\infty$) is assumed in the theoretical calculations. On the 
contrary, for white dwarfs with masses larger than $1.0~M_{\sun}$ the agreement 
is better if we consider a variable star formation rate.  However, for such 
massive stars there is only data for one magnitude bin.  More observations 
corresponding to massive white dwarfs are needed to confirm this behaviour of 
the luminosity function.

For the sake of comparison we have carried out the same calculations but 
considering the expressions of \citet{woo92}.  The resulting luminosity 
functions are shown in Fig.~\ref{fig:Lum-sfrmag2}. In this case, the 
observational data for stars with masses larger than $0.7~M_{\sun}$ is not 
fitted regardless of the star formation rate assumed. On the contrary, the fit 
is better when considering stars with masses above $1.0~M_{\sun}$.  Since there 
are more observational data for the range of masses above $0.7~M_{\sun}$, we 
find more reliable the conclusions that can be obtained from a comparison of 
these data than those obtained from only one magnitude bin, as is the case of 
more massive white dwarfs.  Thus, it seems clear that observations rule out the 
initial-final mass relationship of \citet{woo92}, and that the other 
initial-final mass relationships used in this work seem to be more reliable, 
although the present status of the observational data does not allow to draw a 
definite conclusion about which initial-final mass relationship is more 
adequate.

\section{The white dwarf mass distribution}

\begin{figure*}
\begin{center}
\includegraphics[scale=0.4]{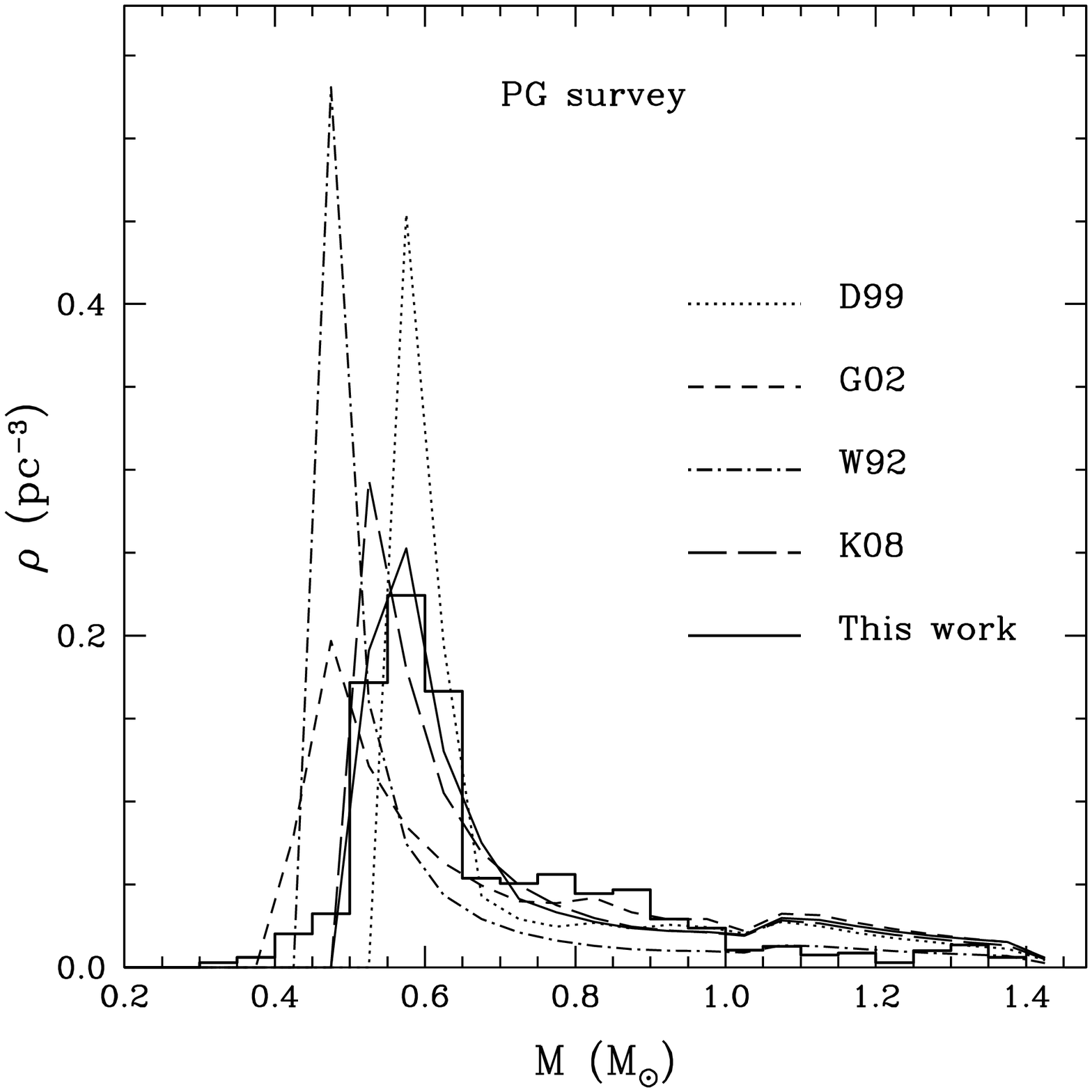}
\includegraphics[scale=0.4]{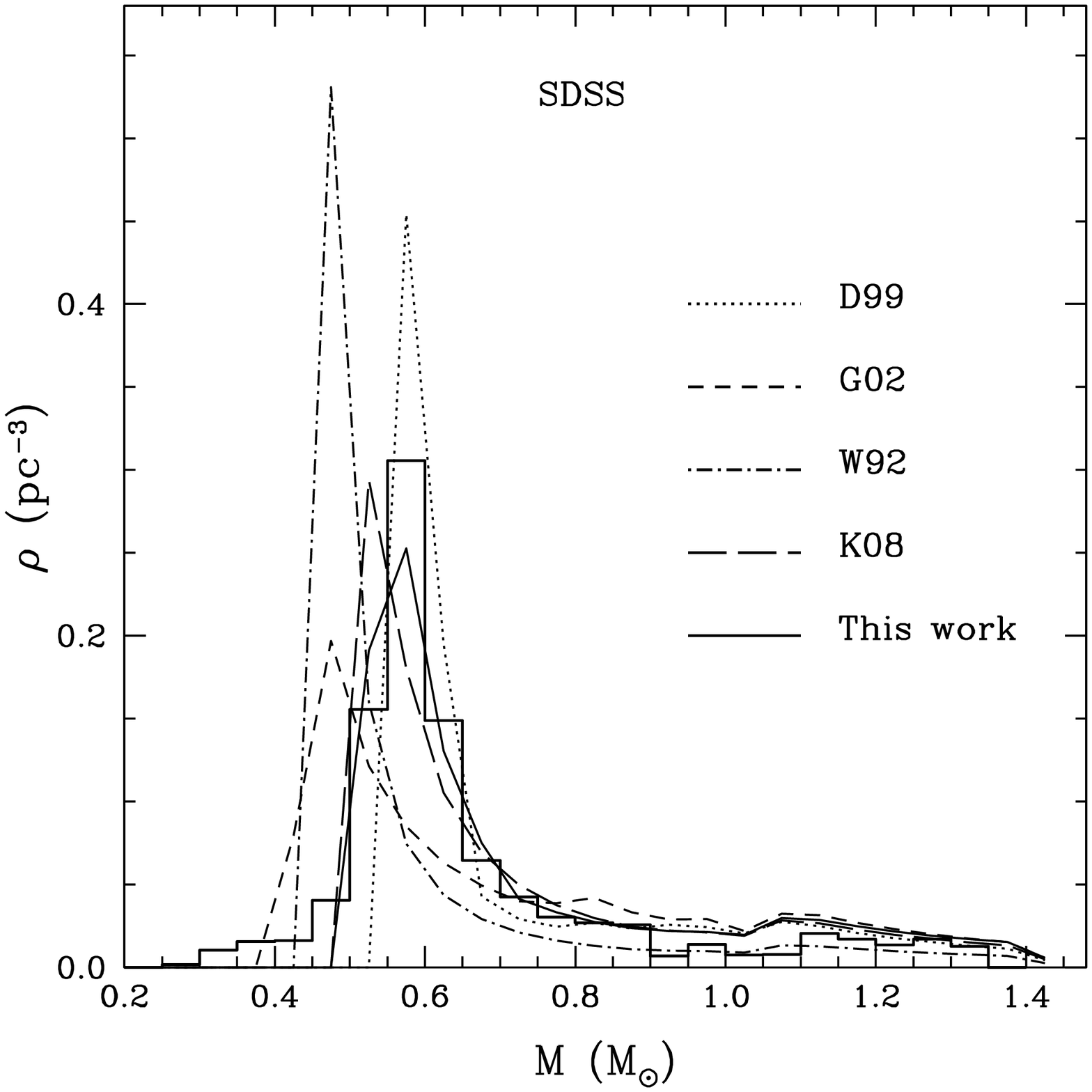}
\caption{Mass   distributions   for    white   dwarfs   with   $T_{\rm
         eff}\geq12\,000$  K considering  different  evolutive stellar
         models  and initial-final  mass  relationships: \citet{dom99}
         --- D99 ---  \citet{gir02} --- G02 ---  \citet{woo92} --- W92
         --- \citet{kal08} ---  K08 ---  and the relation  derived in
         this work.  The histogram  represents the results obtained by
         \citet{lie05b} corresponding to the  data collected in the PG
         Survey   (left)   and   those   obtained   by   \citet{deg08}
         corresponding to the data in the SDSS (right).}
\label{fig:Mass}
\end{center}
\end{figure*}

The understanding of the precise shape of the mass distribution of white dwarfs 
offers a sorely needed insight about the total amount of mass lost during the 
course of the last phases of stellar evolution. Thus, a detailed study of this 
function, from both the theoretical and the observational perspectives can give 
us clues on the initial-final mass relationship \citep{fer05}.  For that 
purpose, we have computed a series of theoretical mass distributions using 
different evolutive stellar models and their corresponding initial-final mass 
relationships, if available.  As in the case of the white dwarf luminosity 
function we have adopted an age of 11 Gyr for the Galactic disc, a constant 
star formation rate and the initial mass function of Salpeter. The theoretical 
mass distributions were then normalized to unit area.

   Our purpose is to compare our results with the observational data obtained 
by \citet{lie05b} from the PG Survey --- left panel of Fig.~\ref{fig:Mass} --- 
and the recent data obtained by \citet{deg08} from the SDSS --- right panel of 
Fig.~\ref{fig:Mass}. As previously pointed out, the accuracy on the mass 
determinations decreases considerable when white dwarfs are cooler than 
$12\,000$ K. Hence, both the theoretical and the observational mass 
distributions in this section consider white dwarfs with $T_{\rm 
eff}\geq12\,000$. In the case of the data from the PG Survey, we have 
computed the observational mass distribution considering the $V_{\rm max}$ 
values reported by \citep{lie05b} taking into account the errors in the masses 
assuming a gaussian distribution. Then, the final distribution has been 
normalized to unit area, as it was done when considering the theoretical mass 
distributions.  In the case of the data from the SDSS, the observational data 
shown in Fig.~\ref{fig:Mass} (right) is the mass distribution computed by 
\citet{deg08}.

In Fig.~\ref{fig:Mass} we plot our results considering the stellar evolutionary 
inputs of \citet{dom99}, \citet{gir02} and \citet{woo92}. We also show our 
results when considering the evolutive stellar models of \citet{dom99} and the 
initial-final mass relation derived in this work, as well as the relation 
recently obtained by \citet{kal08}. All the white dwarf distributions have been 
normalized to the total density obtained in each case. As it can be noted, 
there is a well defined peak in all the mass distributions, the location of 
which is defined mainly by the initial-final mass relationship considered. On 
the contrary, the height of the peak depends also on the lifetime of the 
progenitors.  This can be understood with the help of Fig.~\ref{fig:ifmr}.  
The most abundant stars are those with small masses ($\sim1~M_{\sun}$).  In the 
case of \citet{woo92} and \citet{gir02}, the white dwarfs corresponding to 
these progenitors have masses well below $\sim0.6~M_{\sun}$, and this is the 
reason why the central peak is located at smaller masses in 
Fig.~\ref{fig:Mass}. This peak is then shifted to larger masses when the 
initial-final mass relationship considered favors the production of more 
massive white dwarfs for the low mass progenitors.  This is the case of the 
semi-empirical initial-final mass relationship of \citet{kal08}, our 
relationship or the theoretical relation of \citet{dom99}. Note that in the 
latter case, the production of massive white dwarfs is not favored, opposite to 
what occurs to our semi-empirical relationship, but the peak is located at the 
same mass since low-mass progenitors are the ones that dominate.

If we compare the theoretical mass functions with the observational data of 
\citet{lie05b} from the PG Survey (left panel of Fig.~\ref{fig:Mass})  and the 
recent data obtained by \citet{deg08} from the SDSS (right panel of 
Fig.~\ref{fig:Mass}) it can be noted how in both cases the location of the 
central peak is well fitted by the predictions corresponding to our 
semi-empirical relationship and the theoretical relation of \citet{dom99}. The 
height of the peak is better fitted when considering our semi-empirical 
relationship, although a bit lower in comparison with the SDSS data. The sample 
of white dwarfs corresponding to the SDSS is very complete since it includes 
the 1733 stars with $g\leq19$ and $T_{\rm eff}\geq12\,000$ K present in the 
SDSS DR4, which is seven times the population covered by the PG Survey.

It is worth mentioning that we have not included the He white dwarf population 
in our calculations. This is the reason why there are no white dwarfs below a 
certain mass threshold ($~0.45~M_{\sun}$ in the case of our semi-empirical 
initial-final mass relationship).  The observational data (both from the PG 
survey or the SDSS)  do indeed present a certain number of white dwarfs in this 
low-mass region. On the other hand, we have included the ONe white dwarf 
population which is thought to be dominant for masses larger than 
$~1.05~M_{\sun}$. ONe-core white dwarfs cool faster than CO-core white dwarfs 
due to the large heat capacity of C in comparison with that of O and Ne.  For 
this reason a bump located at $~1.05~M_{\sun}$ in the mass distribution that 
corresponds to the change of the cooling rate is clearly visible.  Since the 
cooling rate is larger for the ONe white dwarf population, there is an increase 
of the number of white dwarfs produced in this region.  As it can be noted in 
Fig.~\ref{fig:Mass} the density of massive white dwarfs is remarkably lower 
when considering the initial-final mass relationship of \citet{woo92} since it 
predicts less massive white dwarfs. In the rest of cases the white dwarf mass 
distributions are approximately coincident in this region.

\section{Summary and Conclusions}

In this work we have revisited the initial-final mass relationship and 
discussed it in a comprehensive manner. With this purpose, we have re-evaluated 
the available data in the literature, mainly based on open clusters, that are 
currently being used to define the semi-empirical initial-final mass 
relationship.  We have used the atmospheric parameters, total ages and 
metallicities reported in the literature and followed the procedure described 
in \cite{cat08} to derive the initial and final masses of these white dwarfs.  
Thanks to these data and our own work based on common proper motion pairs we 
have been able to collect a very heterogeneous sample of white dwarfs, covering 
a wide range of ages and metallicites. Most importantly, with this study we 
have covered the range of initial masses from $1$ to $6.5~M_{\sun}$, which was 
poorly covered until recently \citep{kal08,cat08}.  The extension of the 
initial-final mass relationship to the low-mass end is important since we are 
now studying the most populated region of initial masses according to the 
initial mass function of Salpeter, and this means that we are also reproducing 
the well-established peak of the field white dwarf mass distribution 
\citep{kep07}.

As discussed previously, for each cluster the results present an intrinsic mass 
spread, which may indicate that mass loss could depend more on individual 
stellar properties than on a global mechanism. Thus, mass-loss processes could 
even be a stochastic phenomenon \citep{rei96}, being thus impossible to 
reproduce with models.  Another explanation for the spread of masses found for 
each cluster could lie on the fact that we do not know the internal composition 
of the white dwarfs in our sample, and we are using cooling sequences that have 
fixed values for the C/O ratio at the core and most importantly, fixed 
thicknesses of the H and He envelopes which may not be appropriate to describe 
the cooling of individual white dwarfs in some cases.  This fact could affect 
the cooling times derived in approximately 10 per cent at very low luminosities 
\citep{pra02}.

Since there is no compelling reason to justify the use of a more sophisticated 
relationship, we have performed a weighted least-squares linear fit of these 
data, and from a detailed analysis we have found that there is no correlation 
in this sample between the final masses and metallicity.  We have also given 
some clues on the dependence of the initial-final mass relationship on other 
parameters, such as rotation and magnetism.  In the case of rotation, we have 
probably found the first evidence that rotation may have in stellar evolution, 
corroborating that when the progenitor is a fast rotating star, the resulting 
white dwarf is more massive, in agreement with the study of \citet{dom96}.  
Among our sample there are two magnetic white dwarfs with rather different 
magnetic fields. The masses derived in both cases are also clearly different, 
being more massive the one with stronger magnetic field.  This indicates that 
the intensity of the magnetic field might be related to the mass of the white 
dwarf produced.  However, given the lack of information on other possible 
magnetic white dwarfs belonging to this sample, our results are not conclusive 
enough to assess the impact that magnetic fields may have in the structure of 
white dwarfs.

In the second part of this work we have tested the initial-final mass 
relationship by studying its effect on the luminosity function and mass 
distribution of white dwarfs.  For this purpose we have used different stellar 
evolutionary inputs (stellar tracks and initial-final mass relationships).  We 
have also computed the luminosity function of massive white dwarfs in order to 
evaluate the impact of the initial-final mass relationship. We have noted some 
differences between the theoretical luminosity functions, obtaining a clear 
dependence on the considered initial-final mass relationship, as expected. From 
a comparison of our results with the observational data from the Palomar Green 
Survey we have always obtained a reasonable fit when the range of masses was 
constrained to $M>0.7~M_{\sun}$, except when using the initial-final mass 
relationship of \citet{woo92}. These calculations were performed assuming a 
constant star formation rate, but we have also computed the luminosity 
functions considering an exponentially decreasing star formation rate. We have 
shown that the production of massive white dwarfs is dependent on the assumed 
star formation rate, since in the latter case the density of massive white 
dwarfs drops considerably.  Given the presently available observational data, 
any attempt to discern which initial-final mass relationship better fits the 
data is not feasible.  However, our results favor an initial-final mass 
relationship that produces more massive white dwarfs than the relation of 
\citet{woo92}.  This is an important finding, since the initial-final mass 
relationship of \citet{woo92} is the most commonly used relationship for 
computing the theoretical white dwarf luminosity function.

In the case of the white dwarf mass distribution we have obtained more 
conclusive results.  From a comparison of our results with the observational 
data available from both the PG survey \citep{lie05b} and the SDSS 
\citep{deg08} we have noted that the semi-empirical initial-final mass 
relationship derived in this work is the one that better fits the central peak 
of the mass distribution, being in this way very representative of the white 
dwarf population.  On the contrary, the agreement with the observational data 
is rather poor when the relations of \citet{woo92} and \citet{gir02} are 
considered.

The study carried out in this work evidences the necessity of increasing the 
number of high quality observations of white dwarfs belonging to stellar 
clusters, or common proper motion pairs, or to any system that may allow the 
determination of their total ages and original metallicities with accuracy. On 
the other hand, this increase in the observational data should be accompanied 
by refined theoretical studies of the initial-final mass relationship.

\section*{Acknowledgments}

We thank the referee, D.~Koester for his useful comments and suggestions.
We want to thank S.~DeGennaro and his collaborators for providing their observational 
white dwarf mass distribution based on the SDSS data. S.C. acknowledges
support from the Spanish Ministerio de Educaci\'on y Ciencia (MEC)  through a 
FPU grant. This research was also partially supported by MEC grants 
AYA05--08013--C03--01 and 02, by the European Union FEDER funds and by the 
AGAUR.

\label{lastpage}

\end{document}